\documentclass[nofootinbib,reprint,amsmath,amssymb,prd,aps,superscriptaddress]{revtex4-1}
\usepackage{graphicx}
\usepackage{amsmath}
\usepackage{color}
\usepackage{mathtools}
\usepackage{hyperref}
\usepackage{comment}
\usepackage{mathrsfs}

\begin{document}

\title{From minimal-length quantum theory to modified gravity}

\author{Rocco D'Agostino}
\email{rocco.dagostino@inaf.it}
\affiliation{INAF -- Osservatorio Astronomico di Roma, Via Frascati 33, 00078 Monte Porzio Catone, Italy}
\affiliation{INFN, Sezione di Roma 1, Piazzale Aldo Moro 2, 00185 Roma, Italy}

\author{Pasquale Bosso}
\email{pasquale.bosso@ino.cnr.it}
\affiliation{CNR-INO, Istituto Nazionale di Ottica, Via Campi Flegrei 34,
Pozzuoli, I-80078, Italy}

\author{Giuseppe Gaetano Luciano}
\email{giuseppegaetano.luciano@udl.cat}
\affiliation{Departamento de Qu\'{\i}mica, F\'{\i}sica y Ciencias Ambientales y del Suelo, Escuela Polit\'ecnica Superior -- Lleida, Universidad de Lleida, Av. Jaume II, 69, 25001 Lleida, Spain}

\begin{abstract}

In this work, we consider generalized uncertainty principles (GUPs) that incorporate a minimal length through generic momentum-dependent deformation functions. We thus develop a systematic approach connecting such a framework to effective gravitational actions extending general relativity.
By examining quantum gravity-motivated corrections to black hole entropy induced by the GUP and employing Wald’s formalism, we reconstruct modifications to Einstein's gravity within the contexts of $f(R)$ and $f(R, R_{\mu\nu} R^{\mu\nu})$ theories. 
In this way, we establish a direct mapping between the GUP parameters and the higher-order curvature coefficients in the gravitational Lagrangian. As an illustrative application, we compute corrections to the general relativistic prediction
for light deflection, which in turn allows us to infer a stringent upper bound on the minimal measurable length.
Our results show that GUP-induced effects can be consistently embedded into extended gravity theories, offering a promising framework
for testing quantum gravity phenomenology through astrophysical and cosmological observations.
\end{abstract}

\maketitle

\section{Introduction}

The search for observable imprints of quantum gravity in semi-classical regimes has motivated several phenomenological models. Among them, the generalized uncertainty principle (GUP) has emerged as a prominent framework for encoding Planck-scale corrections to quantum mechanics. By modifying the standard Heisenberg uncertainty relation, the GUP introduces a momentum-dependent deformation implying a minimal measurable length \citep{Kempf:1994su,Nozari:2005mr,Das:2009hs,Pedram:2011xj,Hossenfelder:2012jw,Bosso:2023sxr}. This minimal scale, typically associated with the Planck length, appears in several approaches to quantum gravity, including string theory, loop quantum gravity, and non-commutative geometry~\cite{Gross:Mende:1988,AmelinoCamelia:2002dsr,Rovelli:1998lrr,Maggiore:1993rv,Konishi:1989wk,Capozziello:1999wx,Bosso:2024nmn}.

On the gravitational side, classical general relativity (GR) can be extended by including higher-order curvature invariants in the action. Such extensions lead to theories like $f(R)$ gravity \citep{Starobinsky:1980te,Barrow:1988xh,Capozziello:2003tk,Clifton:2005aj,Starobinsky:2007hu,Sotiriou:2008rp,DeFelice:2010aj,Nojiri:2017ncd,Capozziello:2019cav,DAgostino:2020dhv,Bajardi:2022ocw}, or more generally to Lagrangians depending on both the Ricci scalar and tensor invariants, as in $f(R, R_{\mu\nu} R^{\mu\nu})$ gravity \citep{Carroll:2004de,Bogdanos:2009tn,Bajardi:2023shq}. Such terms arise naturally in effective field theory descriptions of gravity coupled to quantum matter fields \citep{Birrell:1982ix} and are supported by the renormalization structure of quantum field theory in curved spacetime \cite{Sotiriou:2008rp,Nojiri:2010wj}.

In the absence of a complete and experimentally confirmed theory of quantum gravity, a practical strategy is to adopt phenomenological approaches that can bridge quantum modifications with semi-classical gravitational dynamics. In this view, models such as the GUP provide a useful parametrization of quantum gravitational effects at low energies, suggesting how Planck-scale features might imprint themselves in observable regimes.

Several studies have investigated how GUP affects black hole (BH) physics \cite{Scardigli:1999jh,Buoninfante:2019fwr,Adler:2001vs,Medved:2004yu,Custodio:2003jp,Ong:2025ent}, with particular focus on the entropy-area relation \cite{Majumder:2011wl,DAgostino:2019wko,Bosso:2023fnb,Buoninfante:2020guu,Tawfik:2015kga,Chagoya:2024tqv}. Such corrections typically involve logarithmic and inverse-area terms, which are also predicted by higher-curvature gravity theories.
This formal resemblance suggests a deeper structural connection between the two frameworks and motivates a systematic investigation of their correspondence.
Despite these advances, a general methodology for reconstructing effective gravitational Lagrangians from arbitrary GUP deformations is still lacking. 

In this paper, we propose a general framework to establish a systematic connection between GUP-inspired quantum corrections and extended gravitational actions with higher-curvature terms. Starting from a generic expansion of the GUP deformation function in powers of the momentum uncertainty, we derive the corresponding corrections to BH entropy by interpreting the GUP-modified uncertainty as a regulator near the horizon. These entropy corrections are then translated into modifications of the gravitational action using Wald's entropy formula, first within $f(R)$ and then within the more general $f(R, R_{\mu\nu} R^{\mu\nu})$ class of theories. By inverting the resulting relations, we reconstruct the effective Lagrangian and identify the curvature-dependent terms encoding the underlying quantum uncertainty structure. 
Hence, our approach provides a physically motivated dictionary connecting quantum mechanical deformations to semi-classical gravitational dynamics. Moreover, it represents a natural extension of the results of~\cite{Hammad:2015dka} to the general case of polynomial GUP models with deformation functions of arbitrary degree.

The structure of the paper is as follows. In Sec.~\ref{sec:GUP}, we introduce our GUP model and examine its implications for BH entropy. In Sec.~\ref{sec:Wald}, employing Wald’s formalism, we reconstruct the effective gravitational action that reproduces the GUP-corrected entropy within the frameworks of $f(R)$ and $f(R, R_{\mu\nu} R^{\mu\nu})$ gravity. Section~\ref{sec:Map} establishes an explicit correspondence between the GUP parameters and the coefficients in the curvature expansion of the gravitational Lagrangian, while Sec.~\ref{sec:Astro} is devoted to astrophysical applications of the obtained gravity theory. Finally, in Sec.~\ref{sec:Conc}, we present our conclusions and outline future directions. 

Throughout this work, we adopt natural units, $k_B=\hbar=c=1$, such that the Planck length is $\ell_P=\sqrt{G}$, where $G$ denotes Newton’s gravitational constant.

\section{Generalized uncertainty principle and black hole entropy}
\label{sec:GUP}

The GUP is a phenomenological modification of the Heisenberg uncertainty relation incorporating effects expected from quantum gravity. It typically introduces a minimal measurable length, of the order of Planck's scale, through a deformation of the canonical commutator between position and momentum~\cite{Kempf:1994su}.

A commonly studied class of models involves deformations depending on powers of the momentum uncertainty or, equivalently, on the momentum operator, with the two formulations related through the Robertson–Schrödinger relation. A notable example is the model proposed in Ref.~\cite{Kempf:1994su}, where the commutator is modified as $[x,p]=i(1+\beta p^2)$, with $\beta$ a small deformation parameter associated with the squared Planck scale. This leads to the uncertainty relation $\Delta x \Delta p \geq \frac{1}{2}\left[1+\beta(\Delta p)^2\right]$, implying a non-zero minimal position uncertainty $\Delta x_{\text{min}}=\sqrt{\beta}$. The physical consequences of such a minimal length have been extensively discussed in the literature \cite{Scardigli:2003kr,Jizba:2009qf,Amelino-Camelia:2005zpp,Das:2008kaa,Carr:2015nqa,Scardigli:2018jlm,Anacleto:2015mma,Jizba:2022icu,Tawfik:2015rva,Barca:2023shv,Bosso:2023aht,Segreto:2024gcg}.

Alternative formulations based on non-polynomial deformations have also been proposed, such as $[x,p]=i\,e^{\beta p^2}$ \cite{Nouicer:2007jg} or $[x,p]=i(1-\beta p^2)^{-1}$ \cite{Pedram:2011gw}, which lead to qualitatively different physical implications. The use of even powers of the momentum is mainly motivated by symmetry considerations, since such functions remain invariant under parity transformations $p\to -p$, preserving spatial isotropy. Nevertheless, models including odd powers, such as linear momentum terms \cite{Ali:2011fa}, have also been explored; these explicitly break parity symmetry and may arise in frameworks with intrinsic chirality, anisotropic backgrounds, or non-commutative geometries.

To keep the analysis as general as possible, we adopt a framework in which the deformation is represented by an unspecified analytic function $\Xi(\Delta p)$, which admits a power series expansion, namely
\begin{equation}
\Delta x \, \Delta p \geq \frac{1}{2} \left[ 1 + \Xi (\Delta p) \right],
\label{eq:GUP}
\end{equation}
with
\begin{equation}
\Xi (\Delta p) = \sum_{k=1}^\infty {\ell_P^{k}} c_k (\Delta p)^k,
\label{eq:gExpansion}
\end{equation}
where $c_k$ are dimensionless coefficients and $\ell_{P}$ denotes the Planck length.
This form ensures that higher-order contributions are suppressed by powers of the Planck scale, so that standard quantum mechanics is recovered in the low-energy regime $\Delta p \ll \ell_P^{-1}$. It is straightforward to verify that, in the limit where all coefficients $c_k$ vanish except for $c_2$, the above model reduces to the quadratic GUP commonly employed in the literature.

In order to proceed with analytical calculations, it is customary to assume that inequality~\eqref{eq:GUP} is approximately saturated.
This approximation corresponds to the coherent states of the ordinary theory, i.e., states characterized by a minimal uncertainty product~\cite{Kempf:1994su,Pedram:2012ui,Dey:2012tv,Bosso:2017ndq,Bosso:2021koi,Jizba:2023ygi}. In this way, we obtain an implicit relation for $\Delta p$ in terms of $\Delta x$:
\begin{equation}
    \Delta p = \frac{1}{2 \Delta x} \left[ 1 + \Xi (\Delta p) \right].
    \label{eqn:expansion_Dp_Dp}
\end{equation}
Inverting this relation yields
\begin{equation}
\label{invers}
    \Delta p = \sum_{n=0}^\infty \ell_{P}^{n} \frac{\tilde{c}_n}{(2 \Delta x)^{n+1}}\,,
\end{equation}
with $\tilde{c}_0 = 1$ and 
\begin{equation}
    \tilde{c}_n
    = \sum_{\vec{m} \in P(n)} \frac{n^{\underline{|\vec{m}|-1}}}{\vec{m}!} \vec{c}^{\vec{m}}\,, \quad  n>0\,.
\end{equation}
The definitions of all symbols appearing in the above relation, along with the computational details of its derivation, are provided in \ref{apx:series_coef}.
By way of illustration, we report below the expressions for the first five coefficients:
\begin{subequations}
\begin{align}
   &\tilde{c}_1 = c_1\,, \label{eq:c1_tilde}\\
    &\tilde{c}_2 = c_1^2 + c_2\,, \label{eq:c2_tilde}\\
    &\tilde{c}_3 = c_1^3 + 3 c_1 c_2 + c_3\,, \\
    &\tilde{c}_4 = c_1^4 + 6 c_1^2 c_2 + 2 c_2^2 + 4 c_1 c_3 + c_4\,,\\
    &\tilde{c}_5 = c_1^5 + 10 c_1^3 c_2 + 10 c_1 c_2^2 + 10 c_1^2 c_3 + 5 c_2 c_3 + 5 c_1 c_4 + c_5.
\end{align}
\end{subequations}

\subsection{Implications for BH entropy}
\label{IIA}

BH horizons provide a natural arena to explore quantum gravity effects, in particular the minimal measurable length predicted by the GUP, since processes occurring near them, such as Hawking radiation, are governed by quantum gravity.
In this context, the typical size of quantum fluctuations of matter fields cannot be smaller than the minimal length scale. As a consequence, the absorption or emission of a particle by a BH is constrained by this fundamental limit.

This constraint affects the properties of the horizon area. When a particle is absorbed, the minimal increase in the horizon area $\Delta A$ is related to the particle’s position uncertainty $\Delta x$ \cite{Awad:2014bta,Hammad:2015dka}. Incorporating the GUP-modified uncertainty therefore leads to corrections to the quantization of the horizon area and, consequently, to modifications of the Bekenstein entropy:
\begin{equation}
S = \frac{A}{4G} + S_{\mathrm{corr}}\,,
\label{eq:entropyTotal}
\end{equation}
where $S_{\mathrm{corr}}$ contains sub-leading terms due to quantum gravity effects. On the other hand, starting from BH absorption and emission, with a Gedankenexperiment similar to Heisenberg’s microscope, one can arrive at a GUP formulation~\cite{Maggiore:1993rv}.

As noted above, the heuristic derivation of $S_{\mathrm{corr}}$ is based on estimating the minimal area increase $\Delta A$ due to the absorption of a quantum particle with position uncertainty $\Delta x$ and momentum uncertainty $\Delta p$. 
Near the horizon, the particle energy can be approximated as $E \sim \Delta p$. 
Since the minimal increase in BH area associated with its absorption is $\Delta A_{\text{min}} \sim G E \Delta x$ \cite{Hammad:2015dka}, from Eq.~\eqref{invers} we obtain
\begin{equation}
\Delta A_{\min} =  \lambda G \bigg[1 + \sum_{k=1} G^{k/2}\tilde{c}_k (2\Delta x)^{-k} \bigg],
\label{Amin}
\end{equation}
where we have used $\ell_P^2=G$ and all proportionality factors have been absorbed into the coefficient $\lambda$, which is to be determined. 
Hereafter, unless otherwise specified, the summation index $k$ runs from $1$ to $\infty$.

We remark that the relation $\Delta A_\text{min}\sim G E\Delta x$ should be interpreted as a general semiclassical estimate near the horizon. While the area-radius relation $A=4\pi r_H^2$ is purely geometric and theory-independent, the absorption of a quantum with conserved energy $E$ induces a corresponding variation of the BH mass and horizon radius. In extended gravity theories, this response may differ from the Einstein case through an effective gravitational coupling, while remaining linear in the absorbed energy. Consequently, one generically expects $\Delta A_\text{min}\propto G_{\mathrm{eff}} E\Delta x$, with $\Delta x\sim\mathcal{O}(r_H)$ and a proportionality factor depending on the underlying gravitational dynamics~\cite{Jacobson:1995uq,Padmanabhan:2009vy}. In our case, this factor is absorbed into $\lambda$.

Let us now adopt the usual near-horizon identification $\Delta x  = \zeta\sqrt {A}$ \cite{Hammad:2015dka}, where $\zeta$ is a dimensionless constant.
Therefore, Eq.~\eqref{Amin} can be written as a function of the horizon area:
\begin{equation}
\Delta A_{\min} = \lambda G \bigg( 1 + \sum_k d_k A^{-k/2} \bigg)\,,
\label{eq:deltaAexpansion2}
\end{equation}
where the coefficients $d_k$ are directly related to the deformation parameters $c_k$ through $d_k= G^{k/2}\left(2\zeta\right)^{-k}\tilde{c_k}$.

The entropy is interpreted as counting the number of fundamental area units $\Delta A_{\min}$ that compose the total horizon area:
\begin{equation}
S_\text{GUP} = \int \frac{dA}{\Delta A_{\min}} = \frac{1}{\lambda G} \int \frac{dA}{1 + \sum_k d_k A^{-k/2}}\,.
\end{equation}
Expanding for $\sum_k d_k A^{-k/2} \ll 1$ yields
\begin{equation}
S_\text{GUP} = \frac{1}{\lambda G} \int dA\bigg(1 - \sum_k d_k A^{-k/2}\bigg)\,,
\end{equation}
and integrating term by term, we have 
\begin{equation}
S_{\text {GUP}} = \frac{1}{\lambda G}\bigg(A-2d_1\sqrt{A}- d_2 \ln A+ \sum_{k\geq 3} \frac{2 d_k}{k - 2} A^{1 - k/2}\bigg)\,,
\label{eq:SlogFinal}
\end{equation}
up to an integration constant, which can be set to restore the correct dimensions of the argument of the logarithm. We notice that, regardless of the sign of the coefficients $d_k$, the modified entropy in Eq.~\eqref{eq:SlogFinal} is expected to remain positive, since the corrections induced by the GUP are subdominant compared to the leading-order term $A/(4G)$ (see the discussion below Eq.~\eqref{eq:gExpansion}).

Moreover, one should have $\lambda=4$ to recover the standard Bekenstein entropy in the limit where $d_k=0$, $\forall k$. 
Thus, the final GUP entropy formula takes the form:
\begin{equation}
S_\text{GUP} = \frac{A}{4G} + a_1\sqrt{A}- a_2 \ln A + \sum_{k \geq 3} a_k\, A^{1-k/2} \,,
\label{eq:entropyGUPfinal}
\end{equation}
where, to simplify the notation, we have defined
\begin{equation}
a_k =
\begin{cases}
\dfrac{d_2}{4G}, & \text{if } k = 2\,, \\[2mm]
\dfrac{d_k}{2G(k - 2)}, & \text{if } k \neq 2\,.
\end{cases}
\end{equation}

Equation~\eqref{eq:entropyGUPfinal} shows the generic structure of the entropy once GUP corrections are taken into account. 
In general, however, there is no one-to-one correspondence between the coefficients $a_k$ in $S_{\text{GUP}}$ and the coefficients $c_k$ in Eq.~\eqref{eq:GUP}. 
For instance, the logarithmic correction associated with $a_2$ can arise from both linear and quadratic GUP terms (cf. Eq.~\eqref{eq:c2_tilde}), and similar considerations apply to higher-order corrections.

Nevertheless, if for each $k$ only the leading contributions are retained\footnote{For example, if only $c_1\neq0$, the leading-order term is given by Eq.~\eqref{eq:c1_tilde}, since terms $c_1^n$ with $n>1$ are subdominant and therefore $\tilde c_k=0,\, \forall k>1$.}, a direct correspondence between the GUP expansion and the entropy corrections can be inferred. 
In particular, if $c_k=0$ for all $k\neq1$, the Bekenstein entropy acquires a correction scaling as $\sqrt{A}$, characteristic of an infrared (IR) effect. 
This behavior is consistent with results obtained in linear GUP models~\cite{Majumder:2012rtc,Tawfik:2015kga} and also appears in non-commutative quantization frameworks \cite{Nozari:2008rc}. 
Since such a term becomes subdominant in the ultraviolet (UV) regime, it can be neglected when focusing on Planck-scale physics. 
From the GUP perspective, this provides additional support for models in which the linear term is absent, and the leading deformation is quadratic in $\Delta p$~\cite{Kempf:1994su}.

Conversely, if only $c_2\neq0$, the entropy-area relation receives a logarithmic correction. 
Such a term is widely regarded as a universal feature of UV quantum gravity, although different approaches predict different numerical coefficients~\cite{Carlip:2000nv,Kaul:2000kf,Sen:2012dw,Adler:2001vs,Medved:2004yu}. 
Similar corrections also arise from entanglement entropy calculations in BH spacetimes \cite{Solodukhin:2011gn,Alonso-Serrano:2022qvo}. 
Finally, for $c_k\neq0$ with $k\geq3$, the entropy corrections scale as $A^{1-k/2}$, becoming relevant only in the deep UV regime.

In the next section, we use this framework to reconstruct effective gravitational Lagrangians compatible with these entropy corrections.

\section{Modified gravity and Wald entropy}
\label{sec:Wald}

Modified theories of gravity, in particular those described by a generalized action of the form
\begin{equation}
 \mathcal{S}_g = \frac{1}{16 \pi G} \int d^4x \sqrt{-g} \, f(R)\,,
\label{eq:fRaction}
\end{equation}
represent a natural extension of GR. Here, $g$ is the determinant of the metric tensor, and the Lagrangian density $f(R)$ is a non-linear function of the Ricci scalar $R$, encoding higher-order curvature corrections expected from quantum gravitational or effective field theory considerations \cite{Nojiri:2010wj, Sotiriou:2008rp,DAgostino:2025wgl}.
Such extensions emerge naturally within various theoretical frameworks, including 
quantum field theory formulated on curved spacetime~\cite{Birrell:1982ix}, $\alpha$-corrections models in string theory \cite{Zwiebach:1985uq}, and phenomenological approaches to inflation and dark energy \cite{DeFelice:2010aj}.

To extend the Bekenstein entropy derived in GR, 
Wald formulated a general expression applicable to any diffeomorphism-invariant theory of gravity \cite{Wald:1993nt}. For the class of $f(R)$ theories, the Wald entropy is given by \cite{Iyer:1994ys}
\begin{equation}
S_{\text{Wald}} = \frac{1}{4G} \int_{\mathcal{H}} d^2x \sqrt{h} \, f_R(R)\,,
\label{eq:WaldEntropy}
\end{equation}
where $\mathcal{H}$ denotes the BH horizon cross-section, $h$ is the induced metric on $\mathcal{H}$, and $f_R(R) \equiv \frac{df(R)}{d R}$.
This formula can be understood as a modification of the gravitational coupling constant by a scalar function $f_R(R)$ that depends on the local curvature at the horizon.

In modified gravity theories, such as $f(R)$ models, where static BH solutions may exhibit a non-vanishing Ricci scalar \cite{Multamaki:2006zb,delaCruz-Dombriz:2009pzc}, it is convenient to parametrize the horizon curvature in terms of the geometric scale given by the horizon radius. Following Ref.~\cite{Hammad:2015dka}, we thus assume the scaling relation
\begin{equation}
R_H =\frac{\gamma}{r_H^2} = \frac{4\pi\gamma}{A}\,,
\label{eq:R-A}
\end{equation}
where $\gamma$ is a suitable constant that we set as $\gamma=1/(4\pi)$ for simplicity. 
Relationship \eqref{eq:R-A} allows us to express curvature corrections as functions of the horizon area $A$.

Thus, by focusing on constant curvature solutions, the Wald entropy in $f(R)$ gravity can be expressed as \cite{DAgostino:2024sgm}
\begin{equation}
\label{Wald}
S_{\text{Wald}} = \frac{A}{4G} f_R(R_H)\,,
\end{equation}
where we have used
\begin{equation}
A = \int_{\mathcal{H}}d^2 x\, \sqrt{h} = 4 \pi r_H^2\,.
\end{equation}

Using Eq. \eqref{eq:R-A}, and matching Eq.~\eqref{Wald} with the GUP-corrected entropy expression \eqref{eq:entropyGUPfinal},
we find 
\begin{equation}
 f_R(R) = 1 + 4G \bigg[a_1 \sqrt{R}+ a_2 R \ln R + \sum_{k\geq 3} a_k R^{k/2} \bigg].
\label{eq:fR-identified}
\end{equation}
Integrating the last expression, we obtain the corresponding Lagrangian:
\begin{equation}
f= R + 4G \bigg[\frac{2a_1}{3}R^{\frac{3}{2}}+ \frac{a_2 R^2}{4} (\ln R^2 - 1) +\sum_{k\geq 3} \frac{2 a_k}{k+2} R^{1+\frac{k}{2}} \bigg],
\label{eq:fR-final}
\end{equation}
up to an integration constant that may be interpreted as an effective cosmological constant. 
Equation~\ref{eq:fR-final} makes explicit how GUP-induced entropy corrections translate into non-trivial curvature terms in the gravitational action. 
Thus, the quantum mechanical deformation encoded by $\Xi(\Delta p)$ corresponds semi-classically to a modified gravitational Lagrangian.
It is worth reminding that the consistency of this picture requires the positivity of $f_R(R)$ to ensure a well-defined effective gravitational coupling and avoid ghost instabilities~\cite{Sotiriou:2008rp}.

\subsection{Extension to $f(R, R_{\mu\nu} R^{\mu\nu})$ gravity}

While the $f(R)$ gravity framework captures a significant subset of higher-curvature corrections, quantum gravity and effective field theory considerations often necessitate a more general formulation that includes additional curvature invariants, most notably the Ricci tensor squared term $R_{\mu\nu} R^{\mu\nu}$. This corresponds to 
\begin{equation}
\mathcal{S}_g = \frac{1}{16\pi G} \int d^4 x \sqrt{-g} \, f(R, P)\,,
\label{eq:fRPaction}
\end{equation}
where $P \equiv R_{\mu\nu} R^{\mu\nu}$. The presence of the latter terms is well motivated from several perspectives. In quantum field theory in curved spacetime, one-loop effective actions of matter fields generically generate contributions proportional to $R^2$ and $R_{\mu\nu}R^{\mu\nu}$~\cite{Birrell:1982ix,Parker:2009uva}. Such terms also emerge in the renormalization group flow of gravitational couplings and play a key role in asymptotic safety scenarios~\cite{Reuter:1996cp}. Moreover, higher-derivative invariants like $R_{\mu\nu}R^{\mu\nu}$ modify the spectrum of gravitational excitations and can improve the ultraviolet behavior of the theory, potentially leading to perturbative renormalizability~\cite{Stelle:1976gc}.

The Wald entropy formula for a generic action depending on arbitrary curvature invariants reads \cite{DAgostino:2024sgm,DAgostino:2024ymo}
\begin{equation}
S_{\text{Wald}} = -2\pi \int_{\mathcal{H}} d^2 x \sqrt{h} \, \epsilon_{\mu\nu} \epsilon_{\rho\sigma} \frac{\partial \mathcal{L}}{\partial R_{\mu\nu\rho\sigma}}\,,
\end{equation}
where $\epsilon_{\mu\nu}$ is the binormal to the horizon cross-section $\mathcal{H}$, and $\mathcal{L}$ is the Lagrangian density.
For the case $\mathcal{L} = f(R, P)$, one has
\begin{equation}
\frac{\delta \mathcal{L}}{\delta R_{\mu\nu\rho\sigma}} = \sqrt{-g} \left[
f_R\frac{\delta R}{\delta R_{\mu\nu\rho\sigma}} + f_P \frac{\delta P}{\delta R_{\mu\nu\rho\sigma}}
\right],
\end{equation}
where $f_R \equiv \frac{\partial f}{\partial R}$ and $f_P \equiv \frac{\partial f}{\partial P}$.

Taking into account the symmetries of the Riemann tensor and the relative contractions
$R^\mu_{\ \alpha \mu \beta}=R_{\alpha\beta}, R = g^{\alpha\beta}R_{\alpha\beta}$,
one obtains
\begin{equation}
    \frac{\delta R}{\delta R_{\mu\nu\rho\sigma}} = g^{\mu\rho} g^{\nu\sigma} - g^{\mu\sigma} g^{\nu\rho}\,.
\end{equation}
Then, the contraction of the binormal tensors with the metric yields\footnote{Note the antisymmetry of the binormal, $\epsilon_{\mu\nu}=-\epsilon_{\nu\mu}$, and its normalization on the bifurcation surface, $\epsilon_{\mu\nu} \epsilon^{\mu\nu} = -2$.} $\epsilon_{\mu\nu} \epsilon_{\rho\sigma} \left( g^{\mu\rho} g^{\nu\sigma} - g^{\mu\sigma} g^{\nu\rho} \right) = -2$.
Similarly, for $P = R_{\alpha\beta} R^{\alpha\beta}$, we find
\begin{equation}
    \frac{\delta P}{\delta R_{\mu\nu\rho\sigma}} = 2 R^{\alpha\beta} \frac{\delta R_{\alpha\beta}}{\delta R_{\mu\nu\rho\sigma}}\,.
\end{equation}
Since $R_{\alpha\beta} = R^\lambda_{\ \alpha \lambda \beta}$, then 
\begin{equation}
\frac{\delta R_{\alpha\beta}}{\delta R_{\mu\nu\rho\sigma}} = \frac{1}{2} \left(
\delta^\mu_\alpha \delta^\rho_\beta g^{\nu\sigma} - \delta^\mu_\alpha \delta^\sigma_\beta g^{\nu\rho}+\delta^\mu_\beta \delta^\rho_\alpha g^{\nu\sigma} - \delta^\mu_\beta \delta^\sigma_\alpha g^{\nu\rho}\right).
\end{equation}
Hence, one can show that
\begin{equation}
\epsilon_{\mu\nu} \epsilon_{\rho\sigma} \frac{\delta R_{\alpha\beta}}{\delta R_{\mu\nu\rho\sigma}} = -2 n_\alpha n_\beta\,,
\end{equation}
where $n^\mu$ is the unit normal to $\mathcal{H}$.

Using the above results, the Wald entropy reduces to
\begin{equation}
S_{\text{Wald}} = \frac{1}{4G} \int_{\mathcal{H}} d^{2}x\,\sqrt{h} \left[f_R + 2f_P R_{\mu\nu} n^\mu n^\nu
\right].
\label{eq:entropy2}
\end{equation}
The first term, $f_R$, corresponds to the scalar-curvature dependence already present in $f(R)$ gravity. The second term, proportional to $f_P$, encodes the sensitivity of the entropy to local Ricci curvature components normal to the horizon. This term can be interpreted as a correction to the gravitational coupling on the horizon, induced by local anisotropies or tidal effects in the geometry.

For static and spherically symmetric BHs, the curvature scalars on the horizon scale as inverse powers of $r_H$,
\begin{equation}
R \propto \frac{1}{r_H^2}\,, \quad P \propto \frac{1}{r_H^4}\,,
\end{equation}
thus, we can write the following relations: 
\begin{equation}
    R =\frac{1}{A}\,,  \quad R_{\mu\nu}n^\mu n^\nu=\frac{\eta}{A}\,,
    \label{eq:scaling}
\end{equation}
where $\eta$ is a dimensionless constant, with the first of the above relations following from comparison with Eq.~\eqref{eq:R-A}.

Therefore, Eq.~\eqref{eq:entropy2} reads
\begin{equation}
S_\text{Wald} = \frac{A}{4G} f_R\left(R,P\right) + \frac{\eta}{2G} f_P\left(R,P\right).
\label{eq:S_R_P}
\end{equation}
We thus impose the condition
\begin{equation}
\frac{A}{4G} f_R(R,P) + \frac{\eta}{2G} f_P(R,P) = S_{\text{GUP}}(A)\,,
\end{equation}
which can be written as
\begin{equation}
    f_R+\alpha R f_P=h(R)\,,
    \label{eq:PDE}
\end{equation}
where $\alpha\equiv {2\eta}$ and
\begin{equation}
   h(R)\equiv {4GR}\hspace{0.3mm} S_\text{GUP}\,.
    \label{eq:def_h(R)}
\end{equation}
Employing Eq.~\eqref{eq:entropyGUPfinal} with $A=1/R$, we can see that $h(R)$ coincides with Eq.~\eqref{eq:fR-identified}.

The characteristic curves satisfying Eq.~\eqref{eq:PDE} are
\begin{equation}
\frac{dP}{dR} = \alpha R\,\,\, \Longrightarrow\,\, P = \frac{\alpha}{2} R^2 + C\,,
\end{equation}
where $C$ is a constant that identifies the particular characteristic curve. Along each such curve, we have
\begin{equation}
    \frac{df}{dR}=h(R)\,,
\end{equation}
hence, the general solution to Eq.~\eqref{eq:PDE} takes the form
\begin{equation}
f(R,P) = \int h(R) \, dR + \chi(x)\,,
\label{eq:integral_f(R,P)}
\end{equation}
where $\chi(x)$ is an arbitrary function of $x\equiv P - \frac{\alpha}{2} R^2$.
Notice that, for the particular case $\chi(x)=0$, one exactly recovers Eq.~\eqref{eq:fR-final}. 

By defining $F(R)\equiv\int h(R)dR$, Eq.~\eqref{eq:integral_f(R,P)} reads
\begin{equation}
f(R, P) = F(R) + \chi\big(x(R,P)\big).
\end{equation}
Then, from Eq.~\eqref{eq:S_R_P}, we have
\begin{equation}
    S_\text{Wald}=\frac{A}{4G}F'+\left(\frac{\eta}{2G}-\frac{A\alpha R}{4G}\right)\chi'\,,
\end{equation}
where $F'\equiv dF/dR$ and $\chi'\equiv d\chi/dx$.
Now, we note that
\begin{equation}
    \frac{\eta}{2G}-\frac{A\alpha R}{4G}=0\,,
\end{equation}
in view of Eqs.~\eqref{eq:scaling} and \eqref{eq:def_h(R)}.
We can conclude that the function $\chi$ does \emph{not} contribute to the entropy, whose final expression is then given by Eq. \eqref{eq:entropyGUPfinal}.

Our results show that the gravitational Lagrangian $f(R,P)$ cannot be fully determined from the Wald entropy alone. Consequently, $\chi(x)$ remains unconstrained by thermodynamic arguments and must be fixed through additional physical considerations, such as regularity, stability of solutions, cosmological evolution, or consistency with quantum corrections. Although the covariant curvature invariant $P$ does not appear in the entropy expression, it contributes to the gravitational field equations and may affect spacetime dynamics, potentially producing observable effects unrelated to BH thermodynamics.

\section{Coefficient mapping and physical interpretation} 
\label{sec:Map}

The correspondence between entropy corrections induced by the GUP and the structure of the modified gravitational Lagrangian allows for a direct interpretation of higher-curvature contributions in terms of underlying quantum gravity effects. 

\begin{table}
\centering
\begin{tabular}{|c|c|c|}
\hline
GUP index $k$\, & Entropy correction  & GR correction \\
\hline
1 & $\sqrt{A}$ & $R^{3/2}$  \\
2 & $\ln A$ & $R^2\left(\ln R^2-1\right)$  \\
$\geq 3$ & $A^{1-k/2}$ & $R^{1+k/2}$ \\
\hline
\end{tabular}
\caption{Correspondence between GUP modifications and gravity corrections.}
\label{tab:summary}
\end{table}

We summarize our results in Table~\ref{tab:summary}. As discussed in Sec.~\ref{IIA}, when restricting to leading-order corrections in $c_k$, the case $k=1$ yields a contribution proportional to $\sqrt{A}$, corresponding to an $R^{3/2}$ correction in the action, which may indicate IR modifications or non-local effects. The $k=2$ term produces a logarithmic correction to the entropy, associated with the structure $R^2(\ln R^2-1)$ in the action. Higher-order terms with $k \geq 3$ give rise to progressively suppressed entropy corrections, mapping to higher powers of $R$ ($\sim R^{1+k/2}$), and thus become relevant primarily in UV regimes of large curvature.

When considering the more general Lagrangian $f(R,P)$, we demonstrated that the underlying gravity theory includes an undetermined function $\chi(x)$, with $x = P - \frac{\alpha}{2} R^2$, that remains invisible to entropy but can influence spacetime dynamics. 

For the sake of concreteness, we restrict the analysis to GUP corrections up to second order, i.e.
\begin{equation}
   \Xi(\Delta p) = c_1 \ell_P \Delta p + c_2 (\ell_P \Delta p)^2\,,
\end{equation}
which leads to
\begin{equation}
    \Delta p = \frac{1}{2\Delta x} \left[1+\frac{c_1 \ell_P}{2 \Delta x}+\frac{\ell_P^2 (c_1^2+c_2)}{4 \Delta x^2}\right].
\end{equation}
In this case, the entropy reads
\begin{equation}
    S_\text{GUP}=\frac{A}{4G}-\frac{1}{4\zeta}\bigg( c_1\sqrt{\frac{A}{G}}+\frac{c_1^2+c_2}{4\zeta}\ln A\bigg)\,.
    \label{eq:entropy_2}
\end{equation}
Notably, the appearance of both $\sqrt{A}$ and $\ln A$ corrections is in agreement with a wide range of results found in the literature~\cite{Carlip:2000nv, Kaul:2000kf, Sen:2012dw, Adler:2001vs, Medved:2004yu,Nozari:2008rc,Majumder:2012rtc, Tawfik:2015kga}.

From Eq.~\eqref{eq:entropy_2}, we obtain the modified gravity theory
\begin{equation}
    f(R)=R-\frac{2G}{3\zeta}\left[\frac{c_1} {\sqrt{G}} R^{3/2}+\frac{3(c_1^2+c_2)}{32\zeta}R^2\left(1-\ln R^2\right)\right],
    \label{final_f(R)}
\end{equation}
which introduces higher-order and non-polynomial curvature terms. 
The presence of the $R^{3/2}$ and $R^2 \ln R^2$ terms is of particular interest. The former reflects the impact of linear GUP corrections, while the latter combines the effects of both linear and quadratic deformations. Such non-analytic curvature contributions often emerge in quantum corrections to the effective action, e.g., from trace anomalies or conformal field theories on curved backgrounds~\cite{Duff:1993wm,Birrell:1982ix,Avramidi:2000bm}, as well as from gravity models with entropy-based reconstruction schemes~\cite{Cognola:2005de}. This further supports the idea that GUP-inspired corrections may capture essential features of an underlying quantum theory of gravity.

\section{Astrophysical application}
\label{sec:Astro}

As an astrophysical application of the framework developed above, we now consider an extended gravity theory inspired by the Lagrangian~\eqref{final_f(R)}, where we restore the radial dependence of the Ricci scalar $R$.
For this purpose, by varying the action with respect to the metric tensor, we obtain the $f(R)$ field equations \cite{Sotiriou:2008rp}:
\begin{equation}
    R_{\mu\nu} f_R-\frac{1}{2}g_{\mu\nu} f-(\nabla_\mu\nabla_\nu{}-g_{\mu\nu}\Box)f_R=8\pi GT_{\mu\nu}\,,
    \label{eq:FE_f(R)}
\end{equation}
where $\Box\equiv\nabla^\mu\nabla_\mu$ is the D'Alembert operator and $T_{\mu\nu}$ is the energy-momentun tensor of matter fields. These equations reduce to Einstein's ones as soon as $f(R)=R$.
Taking the trace of Eq.~\eqref{eq:FE_f(R)}, one finds
\begin{equation}
   R f_R-2f+3\Box f_R=8\pi G\, T\,.
    \label{eq:trace_f(R}
\end{equation}

In general, solving the $f(R)$ field equations may be highly challenging. 
For the sake of simplicity, we thus consider the quadratic GUP  model, i.e., $c_k=0$ for $k\neq2$. This implies
\begin{equation}
    \Delta x \Delta p = \frac{1}{2}\left(1 + c_2 \ell_P^2 \Delta p^2\right).
    \label{eq:quadratic_GUP}
\end{equation} 
From Eq.~\eqref{final_f(R)} (with $c_1=0$), the above assumption translates to the modified gravity Lagrangian
\begin{equation}
     f(R)=R-\frac{G c_2}{16\zeta^2}R^2\left(1-\ln R^2\right)\approx R-\epsilon R^2\,,
    \label{SimpfR}
\end{equation}
where we have defined 
\begin{equation}
\label{epsilon}
\epsilon \equiv \frac{G c_2}{16\zeta^2}\,.
\end{equation} 
In the last step of Eq.~\eqref{SimpfR}, we have neglected the $R^2\ln R^2$ term, which is sub-leading with respect to the $R^2$ term at astrophysical scales\footnote{We remind that Eq.~\eqref{SimpfR}
is up to an integration constant that appears as $\epsilon R^2\ln\mu$, where $\mu$ has dimensions of $R^{-2}$. In our context, $\mu$ can be fixed such that $\mu R^2\sim1$ at astrophysical scales.}.
This framework corresponds to the Starobinsky model~\cite{Starobinsky:1980te}, which reproduces GR in the limit where the GUP reduces to the Heisenberg principle, namely for $c_2\rightarrow0$.

We can attempt to solve the resulting differential problem in the weak-field slow-motion limit of the theory \cite{Schmidt:1986via,Capozziello:2009zz,Berry:2011pb}. To do that, we search for solutions under the following metric ansatz:
\begin{equation}
    ds^2=(1-2\phi)dt^2-(1+2\psi)(dr^2+r^2d\Omega^2)\,,
\end{equation} 
where $d\Omega^2\equiv d\theta^2+\sin^2\theta\, d\varphi^2$.
In this approach, we can neglect time derivatives and products of the potentials $\phi$ and $\psi$. 
Then, one finds
\begin{equation}
    R=2(2\nabla^2\psi-\nabla^2\phi)\,,
\end{equation}
where $\nabla^2\equiv\delta^{ij}\partial_i\partial_j$.
In addition, we have $T_{\mu\nu}=\delta_\mu^0\delta_\nu^{0}\rho$, 
where $\rho$ is the rest-mass energy density.
Hence, from the (00)-component of Eq.~\eqref{eq:FE_f(R)} and from Eq.~\eqref{eq:trace_f(R}, one finds
\begin{align}
    -\nabla^2\phi-\frac{R}{2}+4\epsilon\nabla^2R&=8\pi G\,,
    \label{eq:00}\\
    -R+12\epsilon\nabla^2 R&=8\pi G\,,
    \label{eq:tr}
\end{align}
respectively. Combining Eqs.~\eqref{eq:00} and $\eqref{eq:tr}$, we obtain
 \begin{equation}
     \nabla^2\phi+2\epsilon\nabla^2R=-4\pi G\rho\,.
 \end{equation}
After some manipulations, we arrive at 
 \begin{equation}
     (1-12\epsilon\nabla^2)\nabla^2\phi=-4\pi G(1-16\epsilon\nabla^2)\rho\,.
     \label{eq:phi}
 \end{equation}
 
The above equation represents a fourth-order gravity theory motivated by quantum considerations. Therefore, the solutions must guarantee a Newtonian behavior in the long-range regime, i.e., they should be exponentially vanishing \cite{Stelle:1977ry}.
In particular, taking $\rho$ as a delta source in Eq.~\eqref{eq:phi}, i.e. $\rho=M\delta$, we get the Yukawa-like solutions
\begin{subequations}
\begin{align}
\label{phisol}
    \phi(r)&=\frac{GM}{r}\bigg(1+\frac{e^{-r/\sqrt{12\epsilon}}}{3}\bigg)\,, \\
    \psi(r)&=\frac{GM}{r}\bigg(1-\frac{e^{-r/\sqrt{12\epsilon}}}{3}\bigg)\,,
    \label{psisol}
\end{align}
\end{subequations}
with $\epsilon>0$ to avoid oscillating behaviors. In light of Eq.~\eqref{epsilon}, this condition implies $c_2>0$. From Eq.~\eqref{eq:quadratic_GUP}, this is precisely the requirement for the existence of a minimal length~\cite{Kempf:1994su}.
It is straightforward to verify that Eqs.~\eqref{phisol}–\eqref{psisol} reproduce the linearized Schwarzschild metric in the regime $r \gg \sqrt{12\epsilon}$, i.e., at distances well above the Planck scale.

We can now examine the observable consequences of the GUP-corrected metric by analyzing the light deflection by a massive body. To this aim, we follow the treatment of \cite{Weinberg:1972kfs,Scardigli:2014qka} and consider a photon approaching the Sun from very large distances. Then, the photon orbit $\vartheta(r)$ in given by
\begin{equation}
\label{eqn:theta}
    \vartheta(r)=\int_\infty^r dr'\, \mathcal{I}(r')\,,
\end{equation}
where 
\begin{equation}
    \mathcal{I}(r)\equiv\frac{1}{r}\left[\left(\frac{r}{r_0}\right)^2\mathcal{F}(r_0)-\mathcal{F}(r)\right]^{-1/2},
\end{equation}
with $\mathcal{F}(r)=1-2\phi(r)$, and $r_0$ being the minimum distance between the Sun and the photon.
The total deflection angle is then calculated as
\begin{equation}
    \Delta\vartheta=2|\gamma(r_0)|-\pi\,.
    \label{eq:Deltatheta}
\end{equation}

To simplify calculations, we write
\begin{equation}
\label{lightdef}
   \mathcal{I}(r)=\frac{1}{r}\bigg[\left(\frac{r}{r_0}\right)^2-1\bigg]^{-1/2}\left[\mathcal{F}(r_0)+\frac{\mathcal{F}(r_0)-\mathcal{F}(r)}{(r/r_0)^2-1}\right]^{-1/2}.
\end{equation}
We recast the argument of the last squared bracket as
\begin{equation}
\notag
\mathcal{F}(r_0)+\frac{\mathcal{F}(r_0)-\mathcal{F}(r)}{(r/r_0)^2-1}= 1-\delta\,,
\label{Fr0}
\end{equation}
where the contribution
\begin{equation}
    \delta\equiv \frac{2GM\left[r^3 \left(3+e^{-r_0/\sqrt{12\epsilon}}\right)-r_0^3 \left(3+e^{-r/\sqrt{12\epsilon}}\right)\right]}{3 r r_0 \left(r^2-r_0^2\right)}
\end{equation}
is expected to be subdominant compared to unity in the weak field regime, i.e. $GM/r_0\ll 1$. We can thus employ the expansion
$(1-\delta)^{-1/2} \simeq 1 + \delta/2+3\delta^2/8$ and recast Eq.~\eqref{lightdef} as
\begin{equation}
    \mathcal{I}(r)\simeq \mathcal{I}_1(r)+ \mathcal{I}_2(r)+\mathcal{I}_3(r)\,,
\end{equation}
where 
\begin{subequations}
\begin{align}
    &\mathcal{I}_1(r)\equiv \frac{r_0}{r \sqrt{r^2-r_0^2}}\,, \\
    &\mathcal{I}_2(r)\equiv \frac{G M \left[r^3 \left(3+e^{-r_0/\sqrt{12\epsilon}}\right)-r_0^3 \left(3+e^{-r/\sqrt{12\epsilon}}\right)\right]}{3 r^2 \left(r^2-r_0^2\right)^{3/2}}\,, \label{eq:I2}\\
    &\mathcal{I}_3(r)\equiv  \frac{G^2 M^2 \left[r^3 \left(3+e^{-r_0/\sqrt{12\epsilon}}\right)-r_0^3 \left(3+e^{-r/\sqrt{12\epsilon}}\right)\right]^2}{6 r^3r_0 \left(r^2-r_0^2\right)^{5/2}}.
    \label{eq:I3}
\end{align}
\end{subequations}

Then, we have
\begin{equation}
    \int_{\infty}^{r_0}dr\, \mathcal{I}_1(r)=-\frac{\pi}{2}\,,
\end{equation}
while the integrals $\mathcal{I}_2(r)$ and $\mathcal{I}_3(r)$ are not as straightforward.  
Since $\sqrt{\epsilon}/r \ll 1$ throughout the entire range of integration (we recall that $\sqrt{\epsilon} \propto \ell_{P}$), it is evident that, for fixed $\epsilon$, the exponential factors $e^{-r/\sqrt{12\epsilon}}$ in Eqs.~\eqref{eq:I2} and \eqref{eq:I3} become increasingly subdominant as $r$ grows relative to $r_{0}$.
In other words, the only non-negligible contributions of these exponentials to the integral~\eqref{eqn:theta} are expected to come from radii close to the minimal value $r_{0}$.
Hence, in Eqs.~\eqref{eq:I2} and \eqref{eq:I3}, we may approximate $e^{-r/\sqrt{12\epsilon}}\approx e^{-r_0/\sqrt{12\epsilon}}$, which ensures that the limit for $r\to r_0^+$ is finite\footnote{The divergences as $r\to r_0^+$ cancel out due to the opposite signs of the terms in Eqs.~\eqref{eq:I2} and \eqref{eq:I3}.}. 
In this case, we find
\begin{align}
    &\int_{\infty}^{r_0}dr\, \mathcal{I}_2(r)\approx -\frac{GM}{r_0}\bigg(1+\frac{e^{-r_0/\sqrt{12\epsilon}}}{3}\bigg)\,, \\
    &\int_{\infty}^{r_0}dr\, \mathcal{I}_3(r)\approx -\left(\frac{G M}{r_0}\right)^2\bigg(1+\frac{e^{-r_0/\sqrt{12\epsilon}}}{3}\bigg)^2\left(\frac{15 \pi }{8}-2\right).
\end{align}
Therefore, from Eq.~\eqref{eq:Deltatheta}, we finally obtain
\begin{align}
    \Delta\vartheta \approx &\ \frac{4 G M}{r_0}\bigg(1+\frac{e^{-r_0/\sqrt{12\epsilon}}}{3}\bigg)+ \notag \\
    &\left(\frac{4 G M}{r_0}\right)^2\left(\frac{15 \pi }{64}-\frac{1}{4}\right)\bigg(1+\frac{e^{-r_0/\sqrt{12\epsilon}}}{3}\bigg)^2.
    \label{eq:final_deflection}
\end{align}
In the limit $\epsilon\to0^+$, Eq.~\eqref{eq:final_deflection} reduces to
\begin{equation}
\Delta\vartheta\approx \frac{4 G M}{r_0}+\left(\frac{G M}{r_0}\right)^2\left(\frac{15 \pi }{4}-\frac{1}
{4}\right) ,
\label{eq:Schw2}
\end{equation}
which coincides with the Schwarzschild prediction 
arising from the second-order expansion in $GM/r_0\ll 1$ \footnote{In the literature, Eq.~\eqref{eq:Schw2} is often expressed as \cite{Epstein:1980dw,Fischbach:1980su,Richter:1982zz}
\begin{equation}
    \Delta\vartheta \approx \frac{4 G M}{b}+\frac{15 \pi }{4}\left(\frac{G M}{b}\right)^2,
    \label{eq:Schw2_b}
\end{equation}
where $b=r_0/\sqrt{1-2GM/r_0}$ is the impact parameter. One can readily check that Eqs.~\eqref{eq:Schw2} and \eqref{eq:Schw2_b} agree up to second order in $GM/r_0\ll 1$.}.

It is now interesting to compare our theoretical result with the measurements of the deflection of light by the Sun. In the case of a photon just grazing the Sun's surface, one usually has
\begin{equation}
\Delta\vartheta_\odot=\frac{4 G M_\odot}{R_\odot}\left(\frac{1+\gamma}{2}\right),
\label{eq:defl_Sun}
\end{equation}
where $GM_\odot\approx 7.5\times 10^9\ \text{eV}^{-1}$ and $R_\odot\approx 3.5\times 10^{15}\, \text{eV}^{-1}$. Here, $\gamma$ is the standard post-Newtonian parameter, whose value has been measured, e.g., with the very long baseline interferometry (VLBI)  \cite{Lambert:2009xy}:
\begin{equation}
    |\gamma-1|\lesssim 1.6\times 10^{-4}\,.
    \label{VLBI}
\end{equation}
Substituting Eq.~\eqref{VLBI} into Eq.~\eqref{eq:defl_Sun} and comparing with Eq.~\eqref{eq:final_deflection}, we obtain
$\epsilon\lesssim 1.5\times 10^{28}\, \text{eV}^{-2}$.

Finally, a rough estimate of the quadratic GUP correction~\eqref{eq:quadratic_GUP} follows from Eq.~\eqref{epsilon}, from which we get $c_2=16\epsilon \zeta^2G^{-1}$.
Recalling that  $\zeta = \Delta x\, A^{-1/2}$ (see below Eq.~\eqref{Amin}) and assuming $\Delta x \sim \ell_P$, we find $c_2\sim16\epsilon A^{-1}$. Since in our framework $A$ must be interpreted as the horizon area of a solar-mass BH, i.e., $A=4\pi R_{S,\odot}^2$ with $R_{S,\odot}\approx 1.5\times 10^{10}\, \text{eV}^{-1}$, we obtain
\begin{equation}
    c_2\sim\frac{4\epsilon}{\pi R_{S,\odot}^2}\lesssim 8.5 \times 10^7\,.
    \label{c2_bound}
\end{equation}
Since $\Delta x_{\text{min}}=\sqrt{c_2}\ell_P$ \cite{Kempf:1994su}, 
from Eq.~\eqref{eq:quadratic_GUP} we finally get 
\begin{equation}
    \Delta x_\text{min}\lesssim 9.2\times 10^3\ell_P\,,
    \label{final_bound}
\end{equation}
which significantly improves upon previous upper bounds in the literature (see, e.g., \cite{Bosso:2023aht} and references therein).

It is instructive to compare our analysis with that presented in \cite{Dias:2016lkg}. In that work, the GUP is related to extended theories of gravity only at the level of linearized weak-field dynamics. Such a procedure, however, does not lead to a covariant gravitational action nor guarantee compatibility with the thermodynamic identities underlying BH entropy. Furthermore, the most stringent bound reported there, inferred from the electron anomalous magnetic moment, yields $\Delta x_\text{min} \lesssim 4.7 \times 10^{-18}\,\text{m} = 2.9 \times 10^{17}\, \ell_P$, which is approximately 13 orders of magnitude weaker than our result.

\section{Conclusions}
\label{sec:Conc}
In this work, we have developed a framework establishing an explicit connection between GUP-inspired quantum corrections and extended gravitational actions containing higher-curvature terms. Starting from a generic deformation of the Heisenberg uncertainty principle, expressed as a power series in the momentum uncertainty $\Delta p$, we showed that the resulting quantum corrections manifest as modifications to BH entropy. Using Wald's Noether-charge formalism, we then mapped these entropy corrections onto corresponding deformations of the gravitational Lagrangian within $f(R)$ theories.

We further extended the analysis to the $f(R, R_{\mu\nu} R^{\mu\nu})$ framework, motivated by the fact that quantum field theory in curved spacetime and effective action approaches naturally generate curvature invariants beyond the Ricci scalar. The resulting effective Lagrangians therefore provide a more realistic description of gravitational dynamics near Planckian regimes.

Our construction reveals a correspondence between the coefficients characterizing the GUP deformation, those entering the BH entropy corrections, and the parameters governing the effective gravitational action. This hierarchical mapping helps bridge the gap between phenomenological quantum gravity models based on GUPs and semi-classical modified gravity theories encoding quantum corrections at the level of the action.

We also investigated potential astrophysical implications of the resulting models. In particular, the second-order GUP leads to Yukawa-like corrections to the Schwarzschild metric in the weak-field regime. By analyzing the Sun's light deflection and comparing the second-order weak-field expansion of the deflection angle with current observational bounds, we derived a stringent constraint on the GUP parameter.

Looking ahead, several directions naturally extend this work. Investigating BH solutions within our framework and confronting them with observational data, such as gravitational-wave signals or BH images, would provide concrete tests of the proposed scenario. In addition, exploring the cosmological implications of these GUP-inspired modified gravity models may shed light on early-universe dynamics, inflation, or the present accelerated expansion, potentially opening new observational avenues to probe quantum gravity phenomenology and minimal-length scenarios.

\acknowledgments
R.D. acknowledges financial support from Istituto Nazionale di Fisica Nucleare (INFN), Sezione di Roma 1, \textit{esperimento} Euclid.
The research of G.G.L. is supported by the Ayuda Postdoctoral of the University of Lleida (project code: X25027).

\appendix

\section{$\Delta p$ series inversion}
\label{apx:series_coef}

In this appendix, we aim to invert Eq.~\eqref{eqn:expansion_Dp_Dp} in order to find an expansion that would describe $\Delta p$ in terms of powers of $\Delta x$.
Let us thus rewrite Eq.~\eqref{eqn:expansion_Dp_Dp} as
\begin{equation}
    \Delta p
    = \sum_{n=0}^\infty \ell_P^{n} \Delta p_n\,.
    \label{eqn:expansion_Dp_Dp_apx}
\end{equation}
Equating the coefficients of a specific power $n$ of $\ell_{P}$ from the left and right hand sides of Eq.~\eqref{eqn:expansion_Dp_Dp}, using Eq.~\eqref{eqn:expansion_Dp_Dp_apx}, we can write
\begin{equation}
    \Delta p_n
    = \frac{1}{2 \Delta x} \sum_{k=1}^n c_k \sum_{\vec{\mu} \in P^{(k)}(n-k)} \binom{k}{\vec{\mu}} \prod_{j=0}^{n-k} \Delta p_j^{\mu_j}\,,
    \label{eqn:n_ord_eq}
\end{equation}
where $P^{(k)}(n-k)$ is the set of the integer partitions of $n-k$ in $k$ parts, that is the set of $k$ non-negative integers whose sum gives $n-k$, while we use the notation $\vec{\mu}$ to designate the set of integers $\mu_j \in \{0,n-k\}$, corresponding to the multiplicities for such partitions, that is $n-k = 0 \, \mu_0 + 1 \, \mu_1 + \dots + n \, \mu_n$ and $\sum_{j=0}^n \mu_j = k$.
Furthermore, the symbol $\binom{k}{\vec{\mu}}$ stands for the multinomial
\begin{equation}
    \binom{k}{\vec{\mu}}
    = \binom{k}{\mu_0, \mu_1, \dots, \mu_n}
    = \frac{k!}{\mu_0! \mu_1! \dots \mu_n!}\,.
\end{equation}

It is possible to invert the relation above, obtaining
\begin{equation}
    \Delta p_n
    = \frac{1}{(2 \Delta x)^{n+1}}
     \sum_{\vec{m} \in P(n)} \frac{n^{\underline{|\vec{m}| - 1}}}{\vec{m}!} \vec{c}^{\vec{m}},
    \label{eqn:gen_ord}
\end{equation}
where $P(n)$ is the set of integer partitions of $n$, corresponding to the set of any length of positive-definite integers whose sum gives $n$.
Here, we have adopted the following notation:
\begin{enumerate}
\renewcommand{\labelenumi}{(\roman{enumi})}
    \item $\vec{m} \in P(n)$ is the array of positive-definite integers such that $\sum_{j} j \, m_j = n$;
    \item $|\vec{m}| = \sum_{j} m_j$;
    \item $\vec{m}! = \prod_{j} m_j!$;
\item $\vec{c}^{\vec{m}} = \prod_{j} c_j^{m_j}$;
    \item $n^{\underline{m}} = n (n-1) \dots (n-m+1)$ is a falling factorial.
\end{enumerate}

In order to prove Eq.~\eqref{eqn:gen_ord}, let us introduce the following generating functions:
\begin{align}
    \Delta P(t)= \sum_{n=0}^\infty \Delta p_n t^n\,, \quad
    C(t)= \sum_{n=0}^\infty c_n t^n,
\end{align}
with $\Delta p_0 = 1/ (2 \Delta x)$ and $c_0=1$.
Using the expansion as in Eq.~\eqref{eqn:n_ord_eq}, we can show that
\begin{equation}
    \Delta P(t)
    = \frac{1}{2 \Delta x} \left[1 + C \left(t \Delta P(t)\right)\right].
\end{equation}
At this point, it is convenient to introduce $Y(t)= t \Delta P(t)$, so that the relation above can be written as
\begin{equation}
    t= \frac{2 \Delta x \, Y}{1 + C(Y)}\,.
    \label{eqn:t_of_Y}
\end{equation}
Furthermore, we notice that
\begin{equation}
    Y(t)= \frac{1}{2 \Delta x} + \sum_{n=0} \Delta p_n t^{n+1}\,.
    \label{eq:Y(t)}
\end{equation}
Therefore, inverting Eq.~\eqref{eqn:t_of_Y} in light of Eq.~\eqref{eq:Y(t)} would give us the expression for $\Delta p_n$.
To achieve this goal, we employ the Lagrange inversion theorem \cite{Surya26112023}, which in our case reads
\begin{equation}
    \Delta p_n= [t^{n+1}] Y(t)
    = \frac{1}{(n+1) (2 \Delta x)^{n+1}} [Y^{n}] \left(1 + C(Y)\right)^{n+1},
\end{equation}
where we use the notation $[x^n] f(x)$ to denote the coefficient of $x^n$ in $f(x)$.
Thus, we observe that
\begin{equation}
    (1+C(Y))^{n+1}
    = \sum_{N=0}^\infty \sum_{\vec{\nu} \in P^{(n+1)}(N)} \binom{n+1}{\vec{\nu}} Y^N \vec{c}^{\vec{\nu}}.
\end{equation}
At this point, since $|\vec{\nu}| = n+1$ and given that the tuple $\vec{\nu}$ may also contain an element $\nu_0$ corresponding to the multiplicity of 0 in the partition of $N$, we have
\begin{equation}
    \binom{n+1}{\vec{\nu}}
    = \frac{(n+1)^{\underline{|\vec{\nu}|}}}{\vec{\nu}!}
    = (n+1) \frac{n^{\underline{\nu_1+\nu_2+\dots-1}}}{\nu_1! \nu_2! \dots}\,.
\end{equation}
Hence, we conclude
\begin{align}
    \Delta p_n
    &= \frac{1}{(2 \Delta x)^{n+1}} \sum_{\vec{\nu} \in P^{(n+1)}(n)} \frac{n^{\underline{\nu_1+\nu_2+\dots-1}}}{\nu_1! \nu_2! \dots} \vec{c}^{\vec{\nu}} \notag\\
    &= \frac{1}{(2 \Delta x)^{n+1}} \sum_{\vec{m} \in P(n)} \frac{n^{\underline{|\vec{m}|-1}}}{\vec{m}!} \vec{c}^{\vec{m}}\,.
\end{align}

\bibliography{references}

@article{Adler:2001vs,
  author       = {Adler, R. J. and Chen, P. and Santiago, D. I.},
  title        = {The Generalized Uncertainty Principle and Black Hole Remnants},
  journal      = {Gen. Rel. Grav.},
  volume       = {33},
  pages        = {2101},
  year         = {2001},
  doi          = {10.1023/A:1015281430411},
}

@article{Chagoya:2024tqv,
    author = "Chagoya, Javier and D{\'\i}az-Salda{\~n}a, I. and Amante, Mario H. and L{\'o}pez-Dom{\'\i}nguez, J. C. and Sabido, M.",
    title = "{Observational constraints on entropic cosmology}",
    eprint = "2412.11260",
    archivePrefix = "arXiv",
    primaryClass = "gr-qc",
    doi = "10.1016/j.physletb.2025.139415",
    journal = "Phys. Lett. B",
    volume = "864",
    pages = "139415",
    year = "2025"
}

@article{Padmanabhan:2009vy,
    author = "Padmanabhan, T.",
    title = "{Thermodynamical Aspects of Gravity: New insights}",
    eprint = "0911.5004",
    archivePrefix = "arXiv",
    primaryClass = "gr-qc",
    doi = "10.1088/0034-4885/73/4/046901",
    journal = "Rept. Prog. Phys.",
    volume = "73",
    pages = "046901",
    year = "2010"
}

@article{Jacobson:1995uq,
    author = "Jacobson, Ted and Kang, Gungwon and Myers, Robert C.",
    title = "{Increase of black hole entropy in higher curvature gravity}",
    eprint = "gr-qc/9503020",
    archivePrefix = "arXiv",
    reportNumber = "MCGILL-94-45, UMDGR-95-047, MCGILL-94--45, UMDGR--95--047",
    doi = "10.1103/PhysRevD.52.3518",
    journal = "Phys. Rev. D",
    volume = "52",
    pages = "3518--3528",
    year = "1995"
}

@article{Tawfik:2015rva,
    author = "Tawfik, Abdel Nasser and Diab, Abdel Magied",
    title = "{Review on Generalized Uncertainty Principle}",
    eprint = "1509.02436",
    archivePrefix = "arXiv",
    primaryClass = "physics.gen-ph",
    reportNumber = "ECTP-2015-01, WLCAPP-2015-01",
    doi = "10.1088/0034-4885/78/12/126001",
    journal = "Rept. Prog. Phys.",
    volume = "78",
    pages = "126001",
    year = "2015"
}

@article{Buoninfante:2019fwr,
    author = "Buoninfante, Luca and Luciano, Giuseppe Gaetano and Petruzziello, Luciano",
    title = "{Generalized Uncertainty Principle and Corpuscular Gravity}",
    eprint = "1903.01382",
    archivePrefix = "arXiv",
    primaryClass = "gr-qc",
    doi = "10.1140/epjc/s10052-019-7164-y",
    journal = "Eur. Phys. J. C",
    volume = "79",
    number = "8",
    pages = "663",
    year = "2019"
}

@article{Custodio:2003jp,
    author = "Custodio, Paulo Sergio and Horvath, J. E.",
    title = "{The Generalized uncertainty principle, entropy bounds and black hole (non)evaporation in a thermal bath}",
    eprint = "gr-qc/0305022",
    archivePrefix = "arXiv",
    reportNumber = "IAG-03-233",
    doi = "10.1088/0264-9381/20/14/103",
    journal = "Class. Quant. Grav.",
    volume = "20",
    pages = "L197--L203",
    year = "2003"
}

@article{Dey:2012tv,
    author = "Dey, Sanjib and Fring, Andreas",
    title = "{Squeezed coherent states for noncommutative spaces with minimal length uncertainty relations}",
    eprint = "1207.3297",
    archivePrefix = "arXiv",
    primaryClass = "hep-th",
    doi = "10.1103/PhysRevD.86.064038",
    journal = "Phys. Rev. D",
    volume = "86",
    pages = "064038",
    year = "2012"
}

@article{Jizba:2023ygi,
    author = "Jizba, Petr and Lambiase, Gaetano and Luciano, Giuseppe Gaetano and Petruzziello, Luciano",
    title = "{Coherent states for generalized uncertainty relations as Tsallis probability amplitudes: New route to nonextensive thermostatistics}",
    eprint = "2308.12368",
    archivePrefix = "arXiv",
    primaryClass = "gr-qc",
    doi = "10.1103/PhysRevD.108.064024",
    journal = "Phys. Rev. D",
    volume = "108",
    number = "6",
    pages = "064024",
    year = "2023"
}

@article{Pedram:2012ui,
    author = "Pedram, Pouria",
    title = "{Coherent States in Gravitational Quantum Mechanics}",
    eprint = "1204.1524",
    archivePrefix = "arXiv",
    primaryClass = "hep-th",
    doi = "10.1142/S0218271813500041",
    journal = "Int. J. Mod. Phys. D",
    volume = "22",
    pages = "1350004",
    year = "2013"
}

@article{Buoninfante:2020guu,
    author = "Buoninfante, Luca and Luciano, Giuseppe Gaetano and Petruzziello, Luciano and Scardigli, Fabio",
    title = "{Bekenstein bound and uncertainty relations}",
    eprint = "2009.12530",
    archivePrefix = "arXiv",
    primaryClass = "hep-th",
    doi = "10.1016/j.physletb.2021.136818",
    journal = "Phys. Lett. B",
    volume = "824",
    pages = "136818",
    year = "2022"
}

@article{Nozari:2005mr,
    author = "Nozari, Kourosh and Azizi, Tahereh",
    title = "{Some aspects of minimal length quantum mechanics}",
    eprint = "quant-ph/0507018",
    archivePrefix = "arXiv",
    doi = "10.1007/s10714-006-0262-9",
    journal = "Gen. Rel. Grav.",
    volume = "38",
    pages = "735--742",
    year = "2006"
}

@article{Pedram:2011xj,
    author = "Pedram, Pouria and Nozari, Kourosh and Taheri, S. H.",
    title = "{The effects of minimal length and maximal momentum on the transition rate of ultra cold neutrons in gravitational field}",
    eprint = "1103.1015",
    archivePrefix = "arXiv",
    primaryClass = "hep-th",
    doi = "10.1007/JHEP03(2011)093",
    journal = "JHEP",
    volume = "03",
    pages = "093",
    year = "2011"
}

@article{Kaul:2000kf,
    author = "Kaul, Romesh K. and Majumdar, Parthasarathi",
    title = "{Logarithmic correction to the Bekenstein-Hawking entropy}",
    eprint = "gr-qc/0002040",
    archivePrefix = "arXiv",
    doi = "10.1103/PhysRevLett.84.5255",
    journal = "Phys. Rev. Lett.",
    volume = "84",
    pages = "5255--5257",
    year = "2000"
}

@article{Sen:2012dw,
    author = "Sen, Ashoke",
    title = "{Logarithmic Corrections to Schwarzschild and Other Non-extremal Black Hole Entropy in Different Dimensions}",
    eprint = "1205.0971",
    archivePrefix = "arXiv",
    primaryClass = "hep-th",
    doi = "10.1007/JHEP04(2013)156",
    journal = "JHEP",
    volume = "04",
    pages = "156",
    year = "2013"
}

@article{Berry:2011pb,
    author = "Berry, Christopher P. L. and Gair, Jonathan R.",
    title = "{Linearized f(R) Gravity: Gravitational Radiation and Solar System Tests}",
    eprint = "1104.0819",
    archivePrefix = "arXiv",
    primaryClass = "gr-qc",
    doi = "10.1103/PhysRevD.83.104022",
    journal = "Phys. Rev. D",
    volume = "83",
    pages = "104022",
    year = "2011",
    note = "[Erratum: Phys.Rev.D 85, 089906 (2012)]"
}

@article{Schmidt:1986via,
    author = "Schmidt, Hans-Juergen",
    title = "{The Newtonian limit of fourth order gravity}",
    eprint = "gr-qc/0106037",
    archivePrefix = "arXiv",
    reportNumber = "UNIPO-MATH-01-JUN-07",
    journal = "Astron. Nachr.",
    volume = "307",
    pages = "339--340",
    year = "1986"
}

@article{Bosso:2021koi,
    author = "Bosso, Pasquale and Luciano, Giuseppe Gaetano",
    title = "{Generalized uncertainty principle: from the harmonic oscillator to a QFT toy model}",
    eprint = "2109.15259",
    archivePrefix = "arXiv",
    primaryClass = "hep-th",
    doi = "10.1140/epjc/s10052-021-09795-1",
    journal = "Eur. Phys. J. C",
    volume = "81",
    number = "11",
    pages = "982",
    year = "2021"
}

@article{Pedram:2011gw,
    author = "Pedram, Pouria",
    title = "{A Higher Order GUP with Minimal Length Uncertainty and Maximal Momentum}",
    eprint = "1110.2999",
    archivePrefix = "arXiv",
    primaryClass = "hep-th",
    doi = "10.1016/j.physletb.2012.07.005",
    journal = "Phys. Lett. B",
    volume = "714",
    pages = "317--323",
    year = "2012"
}

@article{Nouicer:2007jg,
    author = "Nouicer, K.",
    title = "{Quantum-corrected black hole thermodynamics to all orders in the Planck length}",
    eprint = "0704.1261",
    archivePrefix = "arXiv",
    primaryClass = "gr-qc",
    doi = "10.1016/j.physletb.2006.12.072",
    journal = "Phys. Lett. B",
    volume = "646",
    pages = "63--71",
    year = "2007"
}

@article{Anacleto:2015mma,
    author = "Anacleto, M. A. and Brito, F. A. and Passos, E.",
    title = "{Quantum-corrected self-dual black hole entropy in tunneling formalism with GUP}",
    eprint = "1504.06295",
    archivePrefix = "arXiv",
    primaryClass = "hep-th",
    doi = "10.1016/j.physletb.2015.07.072",
    journal = "Phys. Lett. B",
    volume = "749",
    pages = "181--186",
    year = "2015"
}

@article{Scardigli:2018jlm,
    author = "Scardigli, Fabio and Blasone, Massimo and Luciano, Gaetano and Casadio, Roberto",
    title = "{Modified Unruh effect from Generalized Uncertainty Principle}",
    eprint = "1804.05282",
    archivePrefix = "arXiv",
    primaryClass = "hep-th",
    doi = "10.1140/epjc/s10052-018-6209-y",
    journal = "Eur. Phys. J. C",
    volume = "78",
    number = "9",
    pages = "728",
    year = "2018"
}

@article{Carr:2015nqa,
    author = "Carr, Bernard J. and Mureika, Jonas and Nicolini, Piero",
    title = "{Sub-Planckian black holes and the Generalized Uncertainty Principle}",
    eprint = "1504.07637",
    archivePrefix = "arXiv",
    primaryClass = "gr-qc",
    doi = "10.1007/JHEP07(2015)052",
    journal = "JHEP",
    volume = "07",
    pages = "052",
    year = "2015"
}

@article{Bosso:2023aht,
    author = "Bosso, Pasquale and Luciano, Giuseppe Gaetano and Petruzziello, Luciano and Wagner, Fabian",
    title = "{30 years in: Quo vadis generalized uncertainty principle?}",
    eprint = "2305.16193",
    archivePrefix = "arXiv",
    primaryClass = "gr-qc",
    doi = "10.1088/1361-6382/acf021",
    journal = "Class. Quant. Grav.",
    volume = "40",
    number = "19",
    pages = "195014",
    year = "2023"
}

@article{Bosso:2023sxr,
    author = "Bosso, Pasquale and Petruzziello, Luciano and Wagner, Fabian",
    title = "{Minimal length: A cut-off in disguise?}",
    eprint = "2302.04564",
    archivePrefix = "arXiv",
    primaryClass = "hep-th",
    doi = "10.1103/PhysRevD.107.126009",
    journal = "Phys. Rev. D",
    volume = "107",
    number = "12",
    pages = "126009",
    year = "2023"
}

@article{Hossenfelder:2012jw,
    author = "Hossenfelder, Sabine",
    title = "{Minimal Length Scale Scenarios for Quantum Gravity}",
    eprint = "1203.6191",
    archivePrefix = "arXiv",
    primaryClass = "gr-qc",
    doi = "10.12942/lrr-2013-2",
    journal = "Living Rev. Rel.",
    volume = "16",
    pages = "2",
    year = "2013"
}

@article{Kempf:1994su,
  author       = {Kempf, A. and Mangano, G. and Mann, R. B.},
  title        = {Hilbert Space Representation of the Minimal Length Uncertainty Relation},
  journal      = {Phys. Rev.},
  volume       = {D52},
  pages        = {1108},
  year         = {1995},
  doi          = {10.1103/PhysRevD.52.1108},
}

@book{Birrell:1982ix,
  author       = {Birrell, N. D. and Davies, P. C. W.},
  title        = {Quantum Fields in Curved Space},
  publisher    = {Cambridge University Press},
  year         = {1982},
}

@article{Nojiri:2010wj,
  author       = {Nojiri, S. and Odintsov, S. D.},
  title        = {Unified cosmic history in modified gravity: from $f(R)$ theory to Lorentz non-invariant models},
  journal      = {Phys. Rept.},
  volume       = {505},
  pages        = {59},
  year         = {2011},
  doi          = {10.1016/j.physrep.2011.04.001},
}

@article{Sotiriou:2008rp,
  author       = {Sotiriou, T. P. and Faraoni, V.},
  title        = {$f(R)$ Theories Of Gravity},
  journal      = {Rev. Mod. Phys.},
  volume       = {82},
  pages        = {451},
  year         = {2010}, 
  doi          = {10.1103/RevModPhys.82.451},
}

@article{Scardigli:1999jh,
  author       = {Scardigli, Fabio},
  title        = {Generalized Uncertainty Principle in Quantum Gravity from Micro‑Black Hole Gedanken Experiment},
  journal      = {Phys.\ Lett.\ B},
  volume       = {452},
  pages        = {39},
  year         = {1999},
  doi          = {10.1016/S0370-2693(99)00167-7},
}

@article{Majumder:2011wl,
  author       = {Majumder, Barun},
  title        = {Black hole entropy and the modified uncertainty principle: A heuristic analysis},
  journal      = {Phys.\ Lett.\ B},
  volume       = {703},
  pages        = {402},
  year         = {2011},
  doi          = {10.1016/j.physletb.2011.08.026},
}

@article{Gross:Mende:1988,
  author       = {Gross, David J. and Mende, Paul F.},
  title        = {String theory beyond the Planck scale},
  journal      = {Nucl. Phys. B},
  volume       = {303},
  pages        = {407},
  year         = {1988},
  doi          = {10.1016/0550-3213(88)90390-2},
}

@article{Rovelli:1998lrr,
  author       = {Rovelli, Carlo},
  title        = {Loop Quantum Gravity},
  journal      = {Living Rev. Relativ.},
  volume       = {1},
  pages        = {1},
  year         = {1998},
  doi          = {10.12942/lrr-1998-1},
}

@article{AmelinoCamelia:2002dsr,
  author       = {Amelino--Camelia, Giovanni},
  title        = {Doubly Special Relativity},
  journal      = {Nature},
  volume       = {418},
  pages        = {34},
  year         = {2002},
  note         = {see arXiv:gr-qc/0207049 for background},
}

@article{Solodukhin:2011gn,
    author = "Solodukhin, Sergey N.",
    title = "{Entanglement entropy of black holes}",
    eprint = "1104.3712",
    archivePrefix = "arXiv",
    primaryClass = "hep-th",
    doi = "10.12942/lrr-2011-8",
    journal = "Living Rev. Rel.",
    volume = "14",
    pages = "8",
    year = "2011"
}

@article{Tawfik:2015kga,
    author = "Tawfik, Abdel Nasser and El Dahab, Eiman Abou",
    title = "{Corrections to entropy and thermodynamics of charged black hole using generalized uncertainty principle}",
    eprint = "1501.01286",
    archivePrefix = "arXiv",
    primaryClass = "gr-qc",
    reportNumber = "ECTP-2013-04, ECTP-2013-04",
    doi = "10.1142/S0217751X1550030X",
    journal = "Int. J. Mod. Phys. A",
    volume = "30",
    number = "09",
    pages = "1550030",
    year = "2015"
}

@article{Barca:2023shv,
    author = "Barca, Gabriele and Montani, Giovanni",
    title = "{Non-singular gravitational collapse through modified Heisenberg algebra}",
    eprint = "2309.09767",
    archivePrefix = "arXiv",
    primaryClass = "gr-qc",
    doi = "10.1140/epjc/s10052-024-12564-5",
    journal = "Eur. Phys. J. C",
    volume = "84",
    number = "3",
    pages = "261",
    year = "2024",
    note = "[Erratum: Eur.Phys.J.C 84, 865 (2024)]"
}

@article{Segreto:2024gcg,
    author = "Segreto, Sebastiano and Montani, Giovanni",
    title = "{Dynamics of the Mixmaster universe in a non-commutative generalized uncertainty principle framework}",
    eprint = "2407.20476",
    archivePrefix = "arXiv",
    primaryClass = "gr-qc",
    doi = "10.1088/1475-7516/2025/03/061",
    journal = "JCAP",
    volume = "03",
    pages = "061",
    year = "2025"
}

@article{Majumder:2012rtc,
    author = "Majumder, Barun",
    title = "{Black Hole Entropy with minimal length in Tunneling formalism}",
    eprint = "1212.6591",
    archivePrefix = "arXiv",
    primaryClass = "gr-qc",
    doi = "10.1007/s10714-013-1581-2",
    journal = "Gen. Rel. Grav.",
    volume = "45",
    pages = "2403--2414",
    year = "2013"
}

@article{Alonso-Serrano:2022qvo,
    author = "Alonso-Serrano, Ana and Li{\v{s}}ka, Marek",
    title = "{Emergence of quadratic gravity from entanglement equilibrium}",
    eprint = "2212.03168",
    archivePrefix = "arXiv",
    primaryClass = "gr-qc",
    doi = "10.1103/PhysRevD.108.084057",
    journal = "Phys. Rev. D",
    volume = "108",
    number = "8",
    pages = "084057",
    year = "2023"
}

@article{Carlip:2000nv,
  author       = {Carlip, Steven},
  title        = {Logarithmic Corrections to Black Hole Entropy from the Cardy Formula},
  journal      = {Class. Quant. Grav.},
  volume       = {17},
  pages        = {4175},
  year         = {2000},
  doi          = {10.1088/0264-9381/17/20/302},
}

@article{Medved:2004yu,
  author       = {Medved, A. J. M.},
  title        = {A Brief Commentary on Black Hole Entropy},
  journal      = {Class. Quant. Grav.},
  volume       = {22},
  pages        = {133},
  year         = {2005},
  doi          = {10.1088/0264-9381/22/1/011},
}

@article{Zwiebach:1985uq,
  author       = {Zwiebach, Barton},
  title        = {Curvature Squared Terms and String Theories},
  journal      = {Phys. Lett. B},
  volume       = {156},
  pages        = {315--317},
  year         = {1985},
  doi          = {10.1016/0370-2693(85)91616-8},
}

@article{DeFelice:2010aj,
  author       = {De Felice, Antonio and Tsujikawa, Shinji},
  title        = {f(R) Theories},
  journal      = {Living Rev. Rel.},
  volume       = {13},
  pages        = {3},
  year         = {2010},
  doi          = {10.12942/lrr-2010-3},
  eprint       = {arXiv:1002.4928 [gr-qc]},
}

@book{Wald:1993nt,
  author       = {Wald, Robert M.},
  title        = {Black Hole Entropy is Noether Charge},
  year         = {1993},
  note         = {Published in Phys. Rev. D48 (1993) 3427-3431},
  journal      = {Phys. Rev. D},
  volume       = {48},
  pages        = {3427--3431},
  doi          = {10.1103/PhysRevD.48.R3427},
  eprint       = {gr-qc/9307038},
}

@book{Parker:2009uva,
  author       = {Parker, Leonard and Toms, David J.},
  title        = {Quantum Field Theory in Curved Spacetime: Quantized Fields and Gravity},
  publisher    = {Cambridge University Press},
  year         = {2009},
  doi          = {10.1017/CBO9780511813924},
}

@article{Reuter:1996cp,
  author       = {Reuter, Martin},
  title        = {Nonperturbative Evolution Equation for Quantum Gravity},
  journal      = {Phys.\ Rev.\ D},
  volume       = {57},
  pages        = {971},
  year         = {1998},
  doi          = {10.1103/PhysRevD.57.971},
}

@article{Stelle:1976gc,
  author       = {Stelle, K.~S.},
  title        = {Renormalization of Higher‑Derivative Quantum Gravity},
  journal      = {Phys.\ Rev.\ D},
  volume       = {16},
  pages        = {953},
  year         = {1977},
  doi          = {10.1103/PhysRevD.16.953},
}

@article{Hammad:2015dka,
    author = "Hammad, Fay{\c{c}}al",
    title = "{f(R)-modified gravity, Wald entropy, and the generalized uncertainty principle}",
    eprint = "1508.05126",
    archivePrefix = "arXiv",
    primaryClass = "gr-qc",
    doi = "10.1103/PhysRevD.92.044004",
    journal = "Phys. Rev. D",
    volume = "92",
    pages = "044004",
    year = "2015"
}

@article{Ali:2011fa,
    author = "Ali, Ahmed Farag and Das, Saurya and Vagenas, Elias C.",
    title = "{A proposal for testing Quantum Gravity in the lab}",
    eprint = "1107.3164",
    archivePrefix = "arXiv",
    primaryClass = "hep-th",
    doi = "10.1103/PhysRevD.84.044013",
    journal = "Phys. Rev. D",
    volume = "84",
    pages = "044013",
    year = "2011"
}

@article{Bosso:2017ndq,
    author = "Bosso, Pasquale and Das, Saurya and Mann, Robert B.",
    title = "{Planck scale Corrections to the Harmonic Oscillator, Coherent and Squeezed States}",
    eprint = "1704.08198",
    archivePrefix = "arXiv",
    primaryClass = "gr-qc",
    doi = "10.1103/PhysRevD.96.066008",
    journal = "Phys. Rev. D",
    volume = "96",
    number = "6",
    pages = "066008",
    year = "2017"
}

@article{Jizba:2022icu,
    author = "Jizba, Petr and Lambiase, Gaetano and Luciano, Giuseppe Gaetano and Petruzziello, Luciano",
    title = "{Decoherence limit of quantum systems obeying generalized uncertainty principle: New paradigm for Tsallis thermostatistics}",
    eprint = "2201.07919",
    archivePrefix = "arXiv",
    primaryClass = "hep-th",
    doi = "10.1103/PhysRevD.105.L121501",
    journal = "Phys. Rev. D",
    volume = "105",
    number = "12",
    pages = "L121501",
    year = "2022"
}

@article{Jizba:2009qf,
    author = "Jizba, Petr and Kleinert, Hagen and Scardigli, Fabio",
    title = "{Uncertainty Relation on World Crystal and its Applications to Micro Black Holes}",
    eprint = "0912.2253",
    archivePrefix = "arXiv",
    primaryClass = "hep-th",
    doi = "10.1103/PhysRevD.81.084030",
    journal = "Phys. Rev. D",
    volume = "81",
    pages = "084030",
    year = "2010"
}

@article{Amelino-Camelia:2005zpp,
    author = "Amelino-Camelia, Giovanni and Arzano, Michele and Ling, Yi and Mandanici, Gianluca",
    title = "{Black-hole thermodynamics with modified dispersion relations and generalized uncertainty principles}",
    eprint = "gr-qc/0506110",
    archivePrefix = "arXiv",
    doi = "10.1088/0264-9381/23/7/022",
    journal = "Class. Quant. Grav.",
    volume = "23",
    pages = "2585--2606",
    year = "2006"
}

@article{Scardigli:2003kr,
    author = "Scardigli, Fabio and Casadio, Roberto",
    title = "{Generalized uncertainty principle, extra dimensions and holography}",
    eprint = "hep-th/0307174",
    archivePrefix = "arXiv",
    doi = "10.1088/0264-9381/20/18/305",
    journal = "Class. Quant. Grav.",
    volume = "20",
    pages = "3915--3926",
    year = "2003"
}

@article{Capozziello:1999wx,
    author = "Capozziello, S. and Lambiase, G. and Scarpetta, G.",
    title = "{Generalized uncertainty principle from quantum geometry}",
    eprint = "gr-qc/9910017",
    archivePrefix = "arXiv",
    doi = "10.1023/A:1003634814685",
    journal = "Int. J. Theor. Phys.",
    volume = "39",
    pages = "15--22",
    year = "2000"
}

@article{Das:2008kaa,
    author = "Das, Saurya and Vagenas, Elias C.",
    title = "{Universality of Quantum Gravity Corrections}",
    eprint = "0810.5333",
    archivePrefix = "arXiv",
    primaryClass = "hep-th",
    doi = "10.1103/PhysRevLett.101.221301",
    journal = "Phys. Rev. Lett.",
    volume = "101",
    pages = "221301",
    year = "2008"
}

@article{Konishi:1989wk,
    author = "Konishi, Kenichi and Paffuti, Giampiero and Provero, Paolo",
    title = "{Minimum Physical Length and the Generalized Uncertainty Principle in String Theory}",
    reportNumber = "IFUP-TH-46-89, GEF-TH-89-9",
    doi = "10.1016/0370-2693(90)91927-4",
    journal = "Phys. Lett. B",
    volume = "234",
    pages = "276--284",
    year = "1990"
}

@article{Maggiore:1993rv,
    author = "Maggiore, Michele",
    title = "{A Generalized uncertainty principle in quantum gravity}",
    eprint = "hep-th/9301067",
    archivePrefix = "arXiv",
    reportNumber = "IFUP-TH-3-93",
    doi = "10.1016/0370-2693(93)91401-8",
    journal = "Phys. Lett. B",
    volume = "304",
    pages = "65--69",
    year = "1993"
}

@article{Nozari:2008rc,
  author    = {Nozari, Kourosh and Mehdipour, S. Hamid},
  title     = {Quantum gravity and recovery of information in black hole evaporation},
  journal   = {Class. Quant. Grav.},
  volume    = {25},
  pages     = {175015},
  year      = {2008},
  eprint    = {0801.4074},
  archivePrefix = {arXiv},
  primaryClass = {gr-qc}
}

@article{Cognola:2005de,
    author = "Cognola, Guido and Elizalde, Emilio and Nojiri, Shin'ichi and Odintsov, Sergei D. and Zerbini, Sergio",
    title = "{One-loop f(R) gravity in de Sitter universe}",
    eprint = "hep-th/0501096",
    archivePrefix = "arXiv",
    doi = "10.1088/1475-7516/2005/02/010",
    journal = "JCAP",
    volume = "02",
    pages = "010",
    year = "2005"
}

@article{Awad:2014bta,
    author = "Awad, Adel and Ali, Ahmed Farag",
    title = "{Minimal Length, Friedmann Equations and Maximum Density}",
    eprint = "1404.7825",
    archivePrefix = "arXiv",
    primaryClass = "gr-qc",
    doi = "10.1007/JHEP06(2014)093",
    journal = "JHEP",
    volume = "06",
    pages = "093",
    year = "2014"
}

@article{Duff:1993wm,
  author    = {Duff, M.J.},
  title     = {Twenty years of the Weyl anomaly},
  journal   = {Class. Quant. Grav.},
  volume    = {11},
  pages     = {1387--1404},
  year      = {1994},
  eprint    = {hep-th/9308075}
}

@book{Avramidi:2000bm,
    author = "Avramidi, I. G.",
    title = "{Heat kernel and quantum gravity}",
    doi = "10.1007/3-540-46523-5",
    isbn = "978-3-540-67155-8",
    publisher = "Springer",
    address = "New York",
    volume = "64",
    year = "2000"
}

@article{Capozziello:2009zz,
    author = "Capozziello, S. and Piedipalumbo, E. and Rubano, C. and Scudellaro, P.",
    title = "{Testing an exact f(R)-gravity model at Galactic and local scales}",
    eprint = "0906.5430",
    archivePrefix = "arXiv",
    primaryClass = "gr-qc",
    doi = "10.1051/0004-6361/200911992",
    journal = "Astron. Astrophys.",
    volume = "505",
    pages = "21--28",
    year = "2009"
}

@article{Bosso:2024nmn,
    author = "Bosso, Pasquale",
    title = "{Minimal-length quantum field theory: a first-principle approach}",
    eprint = "2407.13235",
    archivePrefix = "arXiv",
    primaryClass = "gr-qc",
    doi = "10.1140/epjc/s10052-024-13281-9",
    journal = "Eur. Phys. J. C",
    volume = "84",
    number = "9",
    pages = "898",
    year = "2024"
}

@article{Stelle:1977ry,
    author = "Stelle, K. S.",
    title = "{Classical Gravity with Higher Derivatives}",
    reportNumber = "Print-77-0417 (BRANDEIS)",
    doi = "10.1007/BF00760427",
    journal = "Gen. Rel. Grav.",
    volume = "9",
    pages = "353--371",
    year = "1978"
}

@article{Scardigli:2014qka,
    author = "Scardigli, Fabio and Casadio, Roberto",
    title = "{Gravitational tests of the Generalized Uncertainty Principle}",
    eprint = "1407.0113",
    archivePrefix = "arXiv",
    primaryClass = "hep-th",
    doi = "10.1140/epjc/s10052-015-3635-y",
    journal = "Eur. Phys. J. C",
    volume = "75",
    number = "9",
    pages = "425",
    year = "2015"
}

@book{Weinberg:1972kfs,
    author = "Weinberg, Steven",
    title = "{Gravitation and Cosmology}: {Principles and Applications of the General Theory of Relativity}",
    isbn = "978-0-471-92567-5, 978-0-471-92567-5",
    publisher = "John Wiley and Sons",
    address = "New York",
    year = "1972"
}

@article{Starobinsky:1980te,
    author = "Starobinsky, Alexei A.",
    editor = "Khalatnikov, I. M. and Mineev, V. P.",
    title = "{A New Type of Isotropic Cosmological Models Without Singularity}",
    doi = "10.1016/0370-2693(80)90670-X",
    journal = "Phys. Lett. B",
    volume = "91",
    pages = "99--102",
    year = "1980"
}

@article{Fischbach:1980su,
    author = "Fischbach, Ephraim and Freeman, Belvin S.",
    title = "{Second Order Contribution to the Gravitational Deflection of Light}",
    reportNumber = "PRINT-80-0787 (PURDUE)",
    doi = "10.1103/PhysRevD.22.2950",
    journal = "Phys. Rev. D",
    volume = "22",
    pages = "2950",
    year = "1980"
}

@article{Richter:1982zz,
    author = "Richter, Gary W. and Matzner, Richard A.",
    title = "{Second-order contributions to gravitational deflection of light in the parametrized post-Newtonian formalism}",
    doi = "10.1103/PhysRevD.26.1219",
    journal = "Phys. Rev. D",
    volume = "26",
    pages = "1219--1224",
    year = "1982"
}

@article{Epstein:1980dw,
    author = "Epstein, R. and Shapiro, I. I.",
    title = "{Post-post-Newtonian deflection of light by the Sun}",
    doi = "10.1103/PhysRevD.22.2947",
    journal = "Phys. Rev. D",
    volume = "22",
    pages = "2947--2949",
    year = "1980"
}

@article{Lambert:2009xy,
    author = "Lambert, S. B. and Le Poncin-Lafitte, C.",
    title = "{Determination of the relativistic parameter gamma using very long baseline interferometry}",
    eprint = "0903.1615",
    archivePrefix = "arXiv",
    primaryClass = "gr-qc",
    doi = "10.1051/0004-6361/200911714",
    journal = "Astron. Astrophys.",
    volume = "499",
    pages = "331",
    year = "2009"
}

@article{Capozziello:2003tk,
    author = "Capozziello, Salvatore and Carloni, Sante and Troisi, Antonio",
    title = "{Quintessence without scalar fields}",
    eprint = "astro-ph/0303041",
    archivePrefix = "arXiv",
    journal = "Recent Res. Dev. Astron. Astrophys.",
    volume = "1",
    pages = "625",
    year = "2003"
}

@article{Nojiri:2017ncd,
    author = "Nojiri, S. and Odintsov, S. D. and Oikonomou, V. K.",
    title = "{Modified Gravity Theories on a Nutshell: Inflation, Bounce and Late-time Evolution}",
    eprint = "1705.11098",
    archivePrefix = "arXiv",
    primaryClass = "gr-qc",
    reportNumber = "PHYS.REPT.-692-(2017)-1-104, Phys.Rept. 692 (2017) 1-104",
    doi = "10.1016/j.physrep.2017.06.001",
    journal = "Phys. Rept.",
    volume = "692",
    pages = "1--104",
    year = "2017"
}

@article{Capozziello:2019cav,
    author = "Capozziello, Salvatore and D'Agostino, Rocco and Luongo, Orlando",
    title = "{Extended Gravity Cosmography}",
    eprint = "1904.01427",
    archivePrefix = "arXiv",
    primaryClass = "gr-qc",
    doi = "10.1142/S0218271819300167",
    journal = "Int. J. Mod. Phys. D",
    volume = "28",
    number = "10",
    pages = "1930016",
    year = "2019"
}

@article{Bajardi:2023shq,
    author = "Bajardi, Francesco and D'Agostino, Rocco",
    title = "{Corrections to general relativity with higher-order invariants and cosmological applications}",
    doi = "10.1142/S0219887824400061",
    journal = "Int. J. Geom. Meth. Mod. Phys.",
    volume = "21",
    number = "10",
    pages = "2440006",
    year = "2024"
}

@article{Bogdanos:2009tn,
    author = "Bogdanos, Charalampos and Capozziello, Salvatore and De Laurentis, Mariafelicia and Nesseris, Savvas",
    title = "{Massive, massless and ghost modes of gravitational waves from higher-order gravity}",
    eprint = "0911.3094",
    archivePrefix = "arXiv",
    primaryClass = "gr-qc",
    doi = "10.1016/j.astropartphys.2010.08.001",
    journal = "Astropart. Phys.",
    volume = "34",
    pages = "236--244",
    year = "2010"
}

@article{Carroll:2004de,
    author = "Carroll, Sean M. and De Felice, Antonio and Duvvuri, Vikram and Easson, Damien A. and Trodden, Mark and Turner, Michael S.",
    title = "{The Cosmology of generalized modified gravity models}",
    eprint = "astro-ph/0410031",
    archivePrefix = "arXiv",
    reportNumber = "FERMILAB-PUB-04-335-A",
    doi = "10.1103/PhysRevD.71.063513",
    journal = "Phys. Rev. D",
    volume = "71",
    pages = "063513",
    year = "2005"
}

@article{Bajardi:2022ocw,
    author = "Bajardi, Francesco and D'Agostino, Rocco and Benetti, Micol and De Falco, Vittorio and Capozziello, Salvatore",
    title = "{Early and late time cosmology: the f(R) gravity perspective}",
    eprint = "2211.06268",
    archivePrefix = "arXiv",
    primaryClass = "gr-qc",
    doi = "10.1140/epjp/s13360-022-03418-8",
    journal = "Eur. Phys. J. Plus",
    volume = "137",
    number = "11",
    pages = "1239",
    year = "2022"
}

@article{Starobinsky:2007hu,
    author = "Starobinsky, Alexei A.",
    title = "{Disappearing cosmological constant in f(R) gravity}",
    eprint = "0706.2041",
    archivePrefix = "arXiv",
    primaryClass = "astro-ph",
    doi = "10.1134/S0021364007150027",
    journal = "JETP Lett.",
    volume = "86",
    pages = "157--163",
    year = "2007"
}

@article{Clifton:2005aj,
    author = "Clifton, Timothy and Barrow, John D.",
    title = "{The Power of General Relativity}",
    eprint = "gr-qc/0509059",
    archivePrefix = "arXiv",
    reportNumber = "DAMTP-2005-86",
    doi = "10.1103/PhysRevD.72.103005",
    journal = "Phys. Rev. D",
    volume = "72",
    number = "10",
    pages = "103005",
    year = "2005",
    note = "[Erratum: Phys.Rev.D 90, 029902 (2014)]"
}

@article{Barrow:1988xh,
    author = "Barrow, John D. and Cotsakis, S.",
    title = "{Inflation and the Conformal Structure of Higher Order Gravity Theories}",
    doi = "10.1016/0370-2693(88)90110-4",
    journal = "Phys. Lett. B",
    volume = "214",
    pages = "515--518",
    year = "1988"
}

@article{DAgostino:2024sgm,
    author = "D'Agostino, Rocco and Luciano, Giuseppe Gaetano",
    title = "{Lagrangian formulation of the Tsallis entropy}",
    eprint = "2408.13638",
    archivePrefix = "arXiv",
    primaryClass = "gr-qc",
    doi = "10.1016/j.physletb.2024.138987",
    journal = "Phys. Lett. B",
    volume = "857",
    pages = "138987",
    year = "2024"
}

@article{DAgostino:2024ymo,
    author = "D'Agostino, Rocco and Luongo, Orlando and Mancini, Stefano",
    title = "{Geometric and topological corrections to Schwarzschild black hole}",
    eprint = "2403.06819",
    archivePrefix = "arXiv",
    primaryClass = "gr-qc",
    doi = "10.1140/epjc/s10052-024-13440-y",
    journal = "Eur. Phys. J. C",
    volume = "84",
    number = "10",
    pages = "1060",
    year = "2024"
}

@article{Iyer:1994ys,
    author = "Iyer, Vivek and Wald, Robert M.",
    title = "{Some properties of Noether charge and a proposal for dynamical black hole entropy}",
    eprint = "gr-qc/9403028",
    archivePrefix = "arXiv",
    doi = "10.1103/PhysRevD.50.846",
    journal = "Phys. Rev. D",
    volume = "50",
    pages = "846--864",
    year = "1994"
}

@article{Das:2009hs,
    author = "Das, Saurya and Vagenas, Elias C.",
    editor = "MacKenzie, Richard and Dick, Rainer",
    title = "{Phenomenological Implications of the Generalized Uncertainty Principle}",
    eprint = "0901.1768",
    archivePrefix = "arXiv",
    primaryClass = "hep-th",
    doi = "10.1139/P08-105",
    journal = "Can. J. Phys.",
    volume = "87",
    pages = "233--240",
    year = "2009"
}

@article{Bosso:2023fnb,
    author = "Bosso, Pasquale and Obreg{\'o}n, Octavio and Rastgoo, Saeed and Yupanqui, Wilfredo",
    title = "{Black hole interior quantization: a minimal uncertainty approach}",
    eprint = "2310.04600",
    archivePrefix = "arXiv",
    primaryClass = "gr-qc",
    doi = "10.1088/1361-6382/ad4fd7",
    journal = "Class. Quant. Grav.",
    volume = "41",
    number = "13",
    pages = "135011",
    year = "2024"
}

@article{Dias:2016lkg,
    author = "Dias, Marco and Hoff da Silva, Julio M. and Scatena, Eslley",
    title = "{Higher-order theories from the minimal length}",
    eprint = "1605.04650",
    archivePrefix = "arXiv",
    primaryClass = "hep-th",
    doi = "10.1142/S0217751X16500871",
    journal = "Int. J. Mod. Phys. A",
    volume = "31",
    number = "16",
    pages = "1650087",
    year = "2016"
}

@article{Ong:2025ent,
    author = "Ong, Yen Chin",
    title = "{GUP Effective metric without GUP: Implications for the sign of GUP parameter and quantum bounce}",
    eprint = "2505.07972",
    archivePrefix = "arXiv",
    primaryClass = "gr-qc",
    doi = "10.1016/j.physletb.2025.139936",
    journal = "Phys. Lett. B",
    volume = "870",
    pages = "139936",
    year = "2025"
}

@article{Multamaki:2006zb,
    author = "Multamaki, Tuomas and Vilja, Iiro",
    title = "{Spherically symmetric solutions of modified field equations in f(R) theories of gravity}",
    eprint = "astro-ph/0606373",
    archivePrefix = "arXiv",
    doi = "10.1103/PhysRevD.74.064022",
    journal = "Phys. Rev. D",
    volume = "74",
    pages = "064022",
    year = "2006"
}

@article{delaCruz-Dombriz:2009pzc,
    author = "de la Cruz-Dombriz, A. and Dobado, A. and Maroto, A. L.",
    title = "{Black Holes in f(R) theories}",
    eprint = "0907.3872",
    archivePrefix = "arXiv",
    primaryClass = "gr-qc",
    doi = "10.1103/PhysRevD.80.124011",
    journal = "Phys. Rev. D",
    volume = "80",
    pages = "124011",
    year = "2009",
    note = "[Erratum: Phys.Rev.D 83, 029903 (2011)]"
}

@article{DAgostino:2020dhv,
    author = "D'Agostino, Rocco and Nunes, Rafael C.",
    title = "{Measurements of $H_0$ in modified gravity theories: The role of lensed quasars in the late-time Universe}",
    eprint = "2002.06381",
    archivePrefix = "arXiv",
    primaryClass = "astro-ph.CO",
    doi = "10.1103/PhysRevD.101.103505",
    journal = "Phys. Rev. D",
    volume = "101",
    number = "10",
    pages = "103505",
    year = "2020"
}

@article{DAgostino:2019wko,
    author = "D'Agostino, Rocco",
    title = "{Holographic dark energy from nonadditive entropy: cosmological perturbations and observational constraints}",
    eprint = "1903.03836",
    archivePrefix = "arXiv",
    primaryClass = "gr-qc",
    doi = "10.1103/PhysRevD.99.103524",
    journal = "Phys. Rev. D",
    volume = "99",
    number = "10",
    pages = "103524",
    year = "2019"
}

@article{DAgostino:2025wgl,
    author = "D'Agostino, Rocco and De Falco, Vittorio",
    title = "{Black hole solutions in the revised Deser-Woodard nonlocal theory of gravity}",
    eprint = "2502.15460",
    archivePrefix = "arXiv",
    primaryClass = "gr-qc",
    doi = "10.1088/1475-7516/2025/07/046",
    journal = "JCAP",
    volume = "07",
    pages = "046",
    year = "2025"
}

@article{Surya26112023,
author = {Erlang Surya and Lutz Warnke},
title = {Lagrange Inversion Formula by Induction},
journal = {The American Mathematical Monthly},
volume = {130},
number = {10},
pages = {944--948},
year = {2023},
publisher = {Taylor \& Francis},
doi = {10.1080/00029890.2023.2251344},
URL = {https://doi.org/10.1080/00029890.2023.2251344},
}

\end{document}